# Symmetry-guided prediction of magnetic-ordered ground states

Yuhui Li[1,2,†], Sike Zeng[3,1,†], Yutong Yu[1], Renzheng Xiong[1], Yu-Jun Zhao[3], Xiaobing Chen[2,*] and Qihang Liu[1,2,*]

[1]*State Key Laboratory of Quantum Functional Materials, Department of Physics, and Guangdong Basic Research Center of Excellence for Quantum Science, Southern University of Science and Technology (SUSTech), Shenzhen 518055, China.*

[2]*Quantum Science Center of Guangdong–Hong Kong–Macao Greater Bay Area (Guangdong), Shenzhen 518045, China.*

[3]*Department of Physics, South China University of Technology, Guangzhou, China*

†These authors contributed equally to this work.

*Emails: chenxiaobing@quantumsc.cn, liuqh@sustech.edu.cn

**Abstract**

**Given the scarcity of experimentally confirmed magnetic structures, the reliable prediction of magnetic ground states is crucial yet remains a long-sought challenge due to the complex magnetic potential energy landscape. Here, we propose a symmetry-guided framework that systematically generates realistic magnetic configurations without requiring any experimental input or prior assumptions such as propagation vectors. Within a hierarchical symmetry-breaking scenario, we incorporate the recently developed oriented spin space group formalism, which captures symmetry-breaking induced by both magnetic ordering and spin-orbit coupling. By performing nonrelativistic and relativistic first-principles calculations, we further establish the energy ladder of the generated magnetic configurations. Exemplified by three prominent unconventional magnets MnTe, $Mn_3Sn$ and $CoNb_3S_6$, we demonstrate that only a few dozen first-principles calculations are sufficient to identify the ground-state magnetic structure. To demonstrate the universality and robustness of our approach, we conduct large-scale benchmark tests on the MAGNDATA database. Our framework successfully reproduces experimentally reported magnetic geometries for 78% of the surveyed materials, among which 93% have their spin orientations successfully generated when considering SOC. Furthermore, in a large-scale first-principles benchmark involving 305 compounds, 82% of experimentally reported magnetic structures are accurately captured within an energy tolerance of 5 meV per magnetic atom from the computed ground states. Beyond reproducing known magnetic configurations, our framework further predicts a variety of low-energy metastable phases, including altermagnets, spin-orbit magnets, and noncollinear antiferromagnets with spin splitting or geometric Hall effect. Our work establishes a general and efficient route toward large-scale prediction of magnetic structures and unconventional magnetic phases, and offers insight into the microscopic origins of magnetic interactions across diverse material systems.**

## I. Introduction

Magnetic materials have attracted significant research interest owing to their broad applications in fundamental physics and advanced technology, including data storage, magnetic sensors, energy conversion, and quantum computation [1]. The magnetic structure plays a central role in magnetic-ordered materials because it dictates the underlying symmetry and thus various properties such as spin-polarized electronic structures and Berry curvatures. Recently, symmetry analysis and high-throughput calculations have effectively accelerated the discovery of functional magnets, exemplified by the identification of hundreds of magnetic topological materials [2–5] and altermagnets [6–8]. Nevertheless, the pursuit of "best of class" material candidates with desirable properties, as well as domain controllability and high transition temperatures, requires significantly larger material pools than for nonmagnetic materials. A major bottleneck lies in the scarcity of reliable magnetic-structure data, both experimental and computational. In stark contrast to the ~318,000 crystal structures cataloged in the Inorganic Crystal Structure Database (ICSD) [9], only ~2,000 magnetic structures have been reported in the experimental MAGNDATA database [10,11]. This significant disparity primarily stems from the experimental complexity of resolving magnetic structures, which often requires sophisticated, resource-intensive neutron diffraction measurements.

The theoretical prediction of magnetic-ordered ground states prior to experimental determination remains a long-sought capability in magnetic materials design [12]. Yet, the fundamental challenge lies in the immense magnetic configuration space and the underlying complexity of magnetic exchange interactions. To address this problem, current computational strategies generally fall into two distinct categories: energy-driven approaches and symmetry-guided approaches. Energy-driven approaches focus on the numerical parameterization or optimization of the magnetic potential energy landscape to identify the magnetic ground state. One prominent direction is to construct effective spin Hamiltonians, parameterized via energy mapping method [13,14], Green's-function-based linear spin wave theory [15–19] or machine learning [20–23]. These Hamiltonian-based methods, however, either require *a priori* assumptions regarding the explicit spin Hamiltonian form, which is particularly difficult for complex systems, or demand extensive high-throughput calculations across hundreds to thousands of random magnetic configurations to train reliable models. Alternatively, the ground-state propagation vector can be determined by minimizing the total energy through spin-spiral calculations [24,25], but they remain insufficient for resolving the explicit magnetic structure within the primitive cell.

In contrast, symmetry-guided approaches systematically restrict the magnetic configuration space by imposing crystallographic symmetry constraints onto the localized magnetic moments. This strategy is fundamentally driven by the physical principle that local minima on the magnetic potential-energy landscape typically coincide with high-symmetry structures [26]. Prominent methodologies include magnetic representation analysis, which decomposes magnetic structures into the basis vectors of irreducible representations of the space group (SG) to determine symmetry-allowed structures [27–29]. Complementarily, cluster multipole theory establishes a complete orthogonal basis set to expand complex magnetic structures into localized multipolar moments adapted to point group symmetries [30–32]. Nevertheless, these conventional symmetry-guided approaches are inherently restricted by the assumption of coupled lattice and spin degrees of freedom. By treating magnetic moments as axial vectors of lattice space [33,34], they cannot completely capture the symmetries of magnetic configurations. This coupling relation, encoded in magnetic representation analysis and magnetic space groups (MSGs) [35], limits the resolution needed to differentiate distinct magnetic phases, directly resulting in too many undetermined continuous free parameters for sampling magnetic configurations.

Fortunately, spin space groups (SSGs) provide a promising route to resolve this bottleneck by decoupling operations in lattice and spin spaces, thereby completely capturing the magnetic geometry [36–39]. While the concept of SSGs dates back decades, the recent establishment of their classification theory makes the systematic generation of candidate magnetic geometries feasible [40–43]. In SSG, the decoupling of spin and lattice spaces implies that the independent O(3) global rotational freedom in spin space. However, in real magnets, this symmetry is inevitably broken by SOC, lifting the degeneracy down to a specific MSG. To bridge this gap, the recently developed oriented spin space groups (OSSGs) serve as parent groups that explicitly specify the orientation of spin space relative to lattice space [44]. Under this architecture, the MSG naturally emerges as a subgroup of the parent OSSG with spin and lattice spaces completely locked. Consequently, the OSSG framework achieves a simultaneous description of magnetic geometry and SOC effects, provides an exhaustive route to uniquely differentiate distinct magnetic structures and identify local energy minima that are inaccessible within MSG analysis [44,45] (see Appendix A).

In this work, we employ an OSSG-based theoretical framework to systematically generate symmetry-allowed magnetic configurations within the hierarchical symmetry-breaking scenario induced by magnetic ordering and SOC. The overall workflow is illustrated in Fig. 1. Starting from crystallographic parent structures, we first generate symmetry-allowed magnetic geometries at the SSG level, where a magnetic geometry refers to the relative spin arrangement independent of the overall spin orientation. The subsequent SOC-induced symmetry lowering is then incorporated through the constraints on the orientation of spin space relative to lattice space in OSSGs. This two-stage approach is physically grounded in the energetic hierarchy of the magnetic Hamiltonian; while strong isotropic exchange interactions dictate the magnetic geometries, the much weaker SOC-induced magneto-crystalline anisotropy subsequently orients these geometries into specific magnetic structures. Combined with first-principles calculations without and with SOC, this framework enables efficient exploration of local energy minima across vast magnetic configuration spaces and establishes the energetic ordering of candidate magnetic states. To evaluate the performance of our symmetry-guided magnetic-structure generation method, we benchmark the framework on the MAGNDATA database, which reports magnetic structures derived from neutron diffraction measurements [10]. Our framework successfully reproduces the experimentally reported magnetic geometries for 78% of materials, among which 93% have their final magnetic structures successfully generated. Furthermore, we perform high-throughput first-principles calculations for 305 magnetic compounds and find that 82% of experimentally reported magnetic geometries are correctly identified among the DFT lowest-energy states. Finally, we also systematically predict low-energy metastable magnetic phases that exhibit various intriguing properties of unconventional magnetism [46,47].

The paper is organized as follows. In Sec. II, we introduce the methodology for the symmetry-guided generation of magnetic configurations, including the construction of SG-SSG-MSG(OSSG) subgroup chains, the generation of symmetry-allowed magnetic geometries from SSG Wyckoff positions, and the refinement of these geometries into magnetic structures with constrained spin orientations using MSG subgroups and OSSG orientation matrices. We also present MAGNDATA-based statistical analyses used to define practical generation constraints. In Sec. III, we demonstrate representative applications to several unconventional magnetic materials, including the collinear altermagnet MnTe, coplanar and noncoplanar anomalous Hall antiferromagnets $Mn_3Sn$ and $CoNb_3S_6$. In Sec. IV, we benchmark both the magnetic-structure generation framework and the first-principles energy evaluation procedure on the MAGNDATA database. Finally, in Sec. V, we discuss the implications, limitations, and future applications of the present framework.

## II. Workflow

Starting from a nonmagnetic crystal structure, magnetic phase transitions can be understood as a hierarchical process of symmetry lowering. As schematically illustrated in Fig. 1, we treat separately the symmetry-breaking induced by magnetic ordering and by SOC. In Sec. IIA, we construct the group-subgroup chains from crystallographic SGs to SSGs and demonstrate how the subsequent symmetry-breaking pathways from SSGs to MSGs are delineated by OSSGs. In Sec. IIB, we outline how magnetic geometries are systematically enumerated from these group-subgroup chains using the spin Wyckoff position of SSG, and how they are further refined into symmetry-inequivalent spin orientations through OSSG-constrained magnetic structures. Finally, in Sec. IIC, we perform a statistical analysis of experimental magnetic structures in the MAGNDATA database to introduce physically motivated constraints on the magnetic-structure generation framework, rendering high-throughput screening computationally tractable.

### A. SG-SSG-MSG(OSSG) subgroup chains

In the SOC-free nonmagnetic (or paramagnetic) limit, the system exhibits $G_{\mathrm{SG}} \times \mathrm{O}(3)$ symmetry, where $G_{\mathrm{SG}}$ denotes the parent crystallographic SG and $\mathrm{O}(3)$ represents the orthogonal group in spin space. The implemented magnetic order lowers this symmetry to an SSG subgroup $G_{\mathrm{SSG}}$. Each element of $G_{\mathrm{SSG}}$ can expresses as $\{U||R|\boldsymbol{\tau}\}$, where $U$ and $\{R|\boldsymbol{\tau}\}$ denote the spin-space and lattice-space operations, respectively. All lattice-space operations form the effective space group of the SSG, denoted as $G_0$, while all spin-space operations form the spin-space point group $G_S$. Since the magnetic geometry can break the underlying lattice translation or rotation symmetry even in the absence of SOC, $G_0$ constitutes a subgroup of the parent SG $G_{\mathrm{SG}}$, satisfying $G_0 \subseteq G_{\mathrm{SG}}$. Physically, the absence of SOC brings the decoupling between the basis of spin space and lattice space, leaving spin space with $\mathrm{O}(3)$ rotational degrees of freedom. A prototypical example is a collinear ferromagnet, whose isotropic interactions preserve a global $\mathrm{O}(3)$ symmetry that spontaneously breaks into a collinear ferromagnetic SSG once a concrete magnetic ground state is realized.

Upon introducing SOC, the independent spin rotational freedom is lifted by fixing the orientation of spin space relative to lattice space, and the symmetry of this magnetic structure can be described by an OSSG, whose elements take the form $\{U_O||R|\boldsymbol{\tau}\}$, where the spin rotation $U_O$ is uniquely constrained by the orientation of spin space relative to lattice space. Among OSSG operations, the subset where spin and spatial operations are completely locked by sharing identical rotation axes and angles forms the corresponding MSGs, ensuring that their proper rotation parts coincide exactly. Under this constraint, the MSG operations can be written in $\theta\{R|\boldsymbol{\tau}\}$, where $\theta = E$ or $T$ denotes the identity or time-reversal operation. The spatial part forms the effective space group of the MSG, denoted as $H_0$. Since the inclusion of SOC typically induces further symmetry breaking, the spatial parts of these three frameworks establish a hierarchy: $H_0 \subseteq G_0 \subseteq G_{\mathrm{SG}}$.

The construction procedure of SG-SSG-MSG subgroup chains is schematically illustrated in Fig. 2(a). Instead of directly enumerating SSG and MSG subgroups from $G_{\mathrm{SG}} \times \mathrm{O}(3)$ or $G_0 \times \{E, T\}$, we identify the SG subgroups $G_0$ from the crystallographic parent SG $G_{\mathrm{SG}}$. For each $G_0$, the corresponding SSGs are systematically constructed through the decomposition of normal subgroups of $G_0$ and group extension,

$$G_{\mathrm{SSG}} = \cup_i \{g_{s_i}||g_i L_0\} \times G_{\mathrm{SO}}, \tag{1}$$

where $L_0$ denotes the normal subgroup of $G_0$, $g_i$ are coset representatives of the quotient group $G_0/L_0$, $g_{s_i}$ are their homomorphically mapped spin operations, and $G_{\mathrm{SO}}$ is the spin-only group describing the dimension of magnetic geometry. Similarly, candidate MSG subgroups can be generated from the effective SG $H_0$,

$$G_{\mathrm{MSG}} = H_0 \text{ or } F_0 \cup T h_0 F_0, \tag{2}$$

where $F_0$ is the index-2 normal subgroup of $H_0$, and $h_0$ is the coset representative of $H_0/F_0$. Therefore, the enumeration of SSG and MSG subgroups can be reformulated as the construction of subgroup chains of the crystallographic parent group $G_{\mathrm{SG}}$. The subgroup chains of $G_{\mathrm{SG}}$ are generated recursively using maximal-subgroup relations pre-compiled from the MAXSUB program on the Bilbao Crystallographic Server [49]. To exhaustively generate all symmetry-inequivalent magnetic structures, only representatives from each conjugate class of SSG and MSG subgroups are retained, as detailed in Appendix B.

These candidate MSGs generated by Eq. (2) define the search space for constructing the final SSG-MSG subgroup chains. A candidate MSG subgroup is compatible with a given SSG subgroup if a global spin-space orientation $Q \in O(3)$ can map the spin rotation $U$ of the SSG onto the lattice rotation of the MSG, requiring that

$$Q^{-1}UQ = \eta(\theta)\det(R)\,R, \tag{3}$$

for all spatially matched SSG and MSG operation pairs ($\{U||R|\tau\}$, $\theta\{R|\tau\}$), where $\eta = 1$ for $\theta = E$ and $\eta = -1$ for $\theta = T$. Once such an orientation $Q$ is found, the SSG operations satisfying Eq. (3) determine the coupled operations of the corresponding MSG subgroup, which can be regarded as an SOC-induced descendant of the given SSG subgroup. The detailed procedure for determining $Q$ is provided in Appendix C. Finally, the established correspondence within each SSG-MSG subgroup chain is naturally accommodated within the OSSG framework, which uniquely captures the specific orientation induced by SOC. Consequently, OSSGs provide a unified symmetry framework to label the generated magnetic structures.

### B. Generation of magnetic structures based on spin Wyckoff positions and OSSG orientations

Having established the SG-SSG-MSG(OSSG) group-subgroup chains, we next use them to generate explicit magnetic structures. The SSG subgroup provides the symmetry constraints for generating magnetic geometries, whereas the MSG subgroup and associated OSSG govern the orientation of spin space relative to lattice space, thereby yielding a definite magnetic structure. In this section, we show how SSG Wyckoff positions are used to assign magnetic moments to crystallographic orbits, thereby enumerating all symmetry-allowed magnetic geometries. These geometries are subsequently resolved into symmetry-inequivalent magnetic structures by incorporating the compatible MSG subgroups obtained in Sec. II A.

The spin Wyckoff positions of SSGs impose symmetry constraints on both atomic coordinates and local magnetic moments [40]. Since spin-space operations act exclusively on the spin degrees of freedom and leave spatial coordinates invariant, the Wyckoff letters and the classification of inequivalent crystallographic orbits are directly inherited from $G_0$. The distinctive characteristic of an SSG Wyckoff position lies in its encoding of the symmetry constraints acting on local magnetic moments. Meanwhile, the multiplicity of the spin Wyckoff position of an SSG may differ from that of the corresponding crystallographic Wyckoff position of $G_0$, when the primitive translation period is modulated by magnetic ordering, resulting in the multiplication of the site multiplicity. Such a flexible cell expansion underscores why the SSG framework provides a more complete description of magnetic propagation compared to the MSG framework, which can only differentiate between odd and even modulation periods.

To enable the systematic generation of magnetic geometries, we construct a standardized database of spin Wyckoff positions based on the spin site-symmetry groups and coset decompositions of SSGs (see details in Appendix D). This database has been compiled and organized in our online database: https://database.findspingroup.com [48]. Once the SSG Wyckoff position is available, the magnetic moments can be directly assigned to the atomic sites of a given crystal structure to generate SSG symmetry-constrained magnetic geometry. An illustrative example is shown in Fig. 2(b). For a parent structure with space group $Pm\bar{3}m$ where magnetic atoms occupy the crystallographic Wyckoff position $3c$, the spin Wyckoff position of $P^{1}m^{3^{1}_{001}}\bar{3}^{2_{100}}m^{m_{001}}1$ directly assigns symmetry-allowed magnetic moments to these $3c$ sites, thereby generating the corresponding SSG-symmetry-allowed magnetic geometry.

The magnetic geometries generated at the SSG level still retain the global spin-rotation freedom of the SOC-free limit. This freedom manifests as a Goldstone-like continuous rotational degeneracy over a flat energy landscape. The introduction of SOC explicitly couples the spin degrees of freedom to the underlying lattice space by constraining their relative orientation. This spin-lattice locking gives rise to magneto-crystalline anisotropy, thereby reshaping the degenerate energy landscape into discrete potential wells. To resolve these geometries into specific magnetic structures, we implement the compatible MSG subgroups and the associated OSSG orientation matrices $Q$ determined in Sec. II A. For each SSG-MSG subgroup chain, the matrix $Q$ defines the locking of the SSG spin frame relative to the lattice frame. The oriented magnetic moments in the lattice frame are then obtained by applying $Q$ to the entire set of SSG-generated moments. As illustrated in Fig. 2(b), the generated magnetic geometry can be mapped into distinct symmetry-inequivalent orientations depending on the OSSG-dictated symmetry-breaking pathway. For example, it can be oriented into two different magnetic structures with MSGs $R\bar{3}m'$ and $R\bar{3}m$, corresponding to OSSGs $P^{1}m^{3^{1}_{111}}\bar{3}^{2_{11-2}}m^{m_{111}}1$ and $P^{1}m^{3^{1}_{111}}\bar{3}^{2_{1-10}}m^{m_{111}}1$, respectively.

In this way, the spin Wyckoff positions first generate SSG symmetry-allowed magnetic geometries for SOC-free calculations, efficiently capturing the dominant isotropic-exchange energy scale. Subsequently, the compatible MSG subgroups and associated OSSG resolve the symmetry-inequivalent magnetic structures stabilized by SOC effects, enabling the evaluation of the weaker magneto-crystalline-anisotropy energy scale.

### C. MAGNDATA statistics and generation constraints

Since the total number of SSGs is infinite and the associate SG-SSG-MSG group-subgroup chains can, in principle, extend indefinitely, an unrestricted enumeration of candidate magnetic structures is not computationally feasible. To balance accuracy and computational efficiency, we introduce several physically motivated bounding constraints for magnetic-structure generation. These constraints are guided by the statistical characteristics of experimentally reported magnetic structures in the MAGNDATA database [10].

As of June 6th, 2026, the MAGNDATA database contains a total of 2241 commensurate magnetic structures [10]. After excluding non-stoichiometric crystal structures, we performed a comprehensive statistical survey on the remaining 1807 stoichiometric magnetic structures. As summarized in Fig. 2(c), the experimentally reported magnetic structures reside within a highly localized and compact region of the symmetry-constrained search space. First, 92.6% of the magnetic structures have an effective space group $G_0$ of SSG with a subgroup-chain level of at most 1 relative to the parent SG $G_{SG}$. Here, the $G_0$ level is defined as the number of maximal-subgroup transitions connecting $G_{SG}$ to $G_0$: level 0 corresponds to $G_0 = G_{SG}$, while level 1 corresponds to maximal subgroups of $G_{SG}$. This indicates that experimentally realized magnetic ordering usually involves only moderate lattice-symmetry reduction, providing statistical support for restricting the search to high-symmetry $G_0$ subgroups in the present framework.

Second, 93.3% of observed magnetic structures exhibit a magnetic-site Wyckoff-position splitting number of at most 1. The splitting number quantifies how a parent crystallographic orbit of the parent $G_{\mathrm{SG}}$ decomposes into symmetry-inequivalent orbits of the SSG. A larger splitting number introduces additional independent magnetic sublattices, thereby leading to a combinatorial explosion of the magnetic-structure space. From an energetic perspective, a larger Wyckoff-position splitting number forces chemically equivalent ions into symmetry-inequivalent local environments. This asymmetry imposes a substantial energy penalty due to the required charge or orbital reorganization, explaining why nature strongly prefers physically equivalent sublattices.

Third, 97.7% of the magnetic structures exhibit a magnetic cell volume expansion factor no greater than four relative to the crystallographic primitive cell. This cut-off is rooted in the fact that magnetic propagation modes often favor high-symmetry points in reciprocal space, whereas low-symmetry wavevectors generally yield incommensurate structures that fall outside our current framework. Limiting the expansion to four elegantly captures three-fold in-plane propagation in triangular and hexagonal lattices, as well as four-fold expansions characteristic of multiple-$q$ magnetic structures.

These statistical results indicate that nature favors the preservation of a substantial portion of the parent crystallographic symmetry during magnetic-order-induced phase transitions. Guided by these data-driven insights, we truncate the generation of magnetic structures by restricting candidates to those satisfying three conditions: the $G_0$ subgroup-chain level of at most 1, the magnetic-order-induced Wyckoff-position splitting number of the magnetic atoms of at most 1, and a magnetic cell volume expansion factor of no greater than four. In addition, we prioritize spin Wyckoff positions of SSG wherein the local site symmetry isolates the scalar magnitude of the local moments as the sole remaining degree of freedom. In these cases, the relative spin arrangement is fixed by the SSG symmetry, while the spin orientations are further constrained by the compatible MSG subgroups and associated OSSG orientation matrices. From a physical standpoint, these configurations correspond to symmetry-enforced local minima on the isotropic energy landscape, where the relative spin angles are locked by symmetry without requiring any energetic relaxation. This strategy frees high-throughput screening from tedious non-collinear angular relaxations. The remaining scalar moment magnitudes are then self-consistently determined by first-principles calculations. These MAGNDATA-guided constraints define a computationally tractable and physically sound search space for high-throughput calculations while comprehensively capturing the most representative experimentally observed magnetic structures.

## III. Examples

Having established the symmetry-guided workflow for predicting magnetic ground states, we now demonstrate its application to three representative unconventional magnets: the noncoplanar anomalous Hall antiferromagnet $CoNb_3S_6$ in Sec. IIIA, the collinear altermagnet MnTe and the coplanar anomalous Hall antiferromagnet $Mn_3Sn$ in Sec. IIIB.

### A. Magnetic ground state of noncoplanar antiferromagnet $CoNb_3S_6$

$CoNb_3S_6$ has recently been reported to host an unconventionally large AHE despite having nearly vanishing net magnetization [50]. However, early neutron scattering experiments reported a collinear antiferromagnetic structure that forbids the AHE [51]. In contrast, more recent neutron experiments reveal that $CoNb_3S_6$ hosts an all-in-all-out-type noncoplanar antiferromagnetic order [52], which permits a nonzero AHE even in the absence of SOC. $CoNb_3S_6$ is formed by Co intercalation in layered hexagonal $NbS_2$. Its nonmagnetic crystallographic space group is $G_{\mathrm{SG}} = P6_322$ (No. 182) with magnetic Co atoms occupying the crystallographic Wyckoff position $2c$. Based on this crystal structure and the generation constraints as described in Sec. IIC, we generate 27 inequivalent magnetic geometries, with only the magnetic moment magnitudes remaining undetermined. A selected portion of the corresponding SG-SSG-MSG subgroup chains is shown in Fig. 3(a), together with the spin Wyckoff positions of magnetic sites and the numbers of inequivalent magnetic geometries under different $G_0$-derived SSG subgroups. For example, the level-0 effective SG ($G_0 = G_{\mathrm{SG}}$) gives rise to 133 SSG subgroups under the magnetic cell expansion cutoff of four, from which 9 inequivalent magnetic geometries are generated.

After performing nonrelativistic DFT calculations for the 27 magnetic configurations, the ground-state magnetic geometry can be identified. The calculated energies are ranked in ascending order, with the ten lowest-energy configurations marked in Fig. 3(b). Some representative magnetic configurations are also illustrated in the Fig. 3(b). The detailed information regarding DFT calculations is given in Appendix E. The calculated ground state, corresponding to the noncoplanar SSG 182.4.4.2 ($P^{3^{2}_{-11-1}}6_3^{m_{110}}2^{m_{011}}2|(2_{001},2_{100},1)$), is consistent with the recently reported all-in-all-out-type noncoplanar magnetic configuration [52], whereas the collinear magnetic configuration reported by Parkin *et al.* [51], labeled by SSG 20.17.2.1.L ($P_C^{1}2^{1}2^{1}2_1^{\infty m}1$), is a metastable state with the total energy only 5.1 meV/mag. atom higher than that of the ground state. Recent studies suggest that although such a collinear magnetic configuration manifests spin degeneracy throughout the Brillouin zone, it exhibits a new type of chiral Dirac-like fermion due to a hidden $SU(2)$ symmetry within the SSG regime [53,54].

We further investigate the physical properties dictated by the SSG symmetry for all 27 magnetic configurations in the absence of SOC, including electric polarization, magnetization, spin polarization, and the AHE. The all-in-all-out-type ground-state magnetic geometry has an effective magnetic point group 62'2', which leaves the $z$ component of the Berry curvature invariant. Consequently, a finite AHE is symmetry-allowed even in the absence of SOC [40,43]. In addition, several coplanar antiferromagnetic configurations exhibit nonrelativistic spin polarization. For example, the third-lowest-energy metastable state, which lies merely 4.5 meV/mag. atom above the ground state, could therefore potentially be realized experimentally. Interestingly, it shows an odd-parity $p$-wave spin texture [55–57], as illustrated in Fig. 3(c), where the only remaining $S_z$ component is permitted by the ${}^{4}1$ $k$-point little cogroup (1 denotes the identity operation in real space, 4 represents a fourfold rotation about the $z$ axis in spin space) at general $k$ points, and the odd-parity spin texture is enforced by the coplanar spin-only operations $\{m_z||E|\mathbf{0}\}$ ($m_z$ is the mirror of $z$ plane).

We next determine the possible magnetic orientations of the ground-state magnetic geometry using the SSG-MSG subgroup chains. For the lowest-energy SSG 182.4.4.2, three MSG subgroups with explicit orientation constraints, labeled by the BNS number 150.27 ($P32'1$), 149.23 ($P312'$) and 20.33 ($C2'2'2_1$), are used to construct four oriented magnetic structures, with the MSG 20.33 ($C2'2'2_1$) giving rise to two distinct spin orientations. These structures are then evaluated by SOC-included first-principles calculations to determine the energetically preferred magnetic orientation, as shown in Fig. 3(d). The calculated ground-state magnetic orientation with $P32'1$ symmetry also yields good agreement with the experimental observations [52]. Under SOC, the single Co spin Wyckoff orbit in the SSG symmetry splits into two inequivalent magnetic orbits, allowing a weak magnetic moment of 0.03 $\mu_B$/Co along the crystallographic $c$ axis. This SOC-induced net magnetization is a characteristic feature of spin-orbit magnetism (SOM) [44]. Therefore, this two-stage SSG-MSG strategy separates the dominant isotropic-exchange-driven magnetic geometry from SOC-induced secondary effects such as weak magnetization, providing a symmetry-based route to analyze the microscopic origin of magnetism.

### B. Magnetic ground states of altermagnet MnTe and coplanar antiferromagnet $Mn_3Sn$

We further apply our approach to two additional unconventional magnets: the collinear altermagnet MnTe [6,58,59] and the coplanar anomalous Hall antiferromagnet $Mn_3Sn$ [60]. Both compounds crystallize in the same space group, $P6_3/mmc$ (No. 194), but the magnetic Mn atoms occupy different crystallographic Wyckoff positions, $2a$ in MnTe and $6h$ in $Mn_3Sn$. Using the SG-SSG subgroup chains of the parent space group $P6_3/mmc$, a total of 33 and 48 inequivalent magnetic geometries is generated for MnTe and $Mn_3Sn$, respectively, under the generation constraints described in Sec. IIC. To determine their ground magnetic geometries, we further perform nonrelativistic DFT calculations. Figs. 4(a) and 4(b) show the ten lowest-energy configurations of MnTe and $Mn_3Sn$, respectively.

Our calculated ground geometry of MnTe is characterized by a collinear interlayer antiferromagnetic order and the SSG 194.164.1.1.L ($P^{-1}6_3/^{-1}m^{1}m^{-1}c^{\infty m}1$), while that of $Mn_3Sn$ is characterized by a coplanar intralayer triangular antiferromagnetic order and the SSG 194.11.1.1.P ($P^{3^{2}_{001}}6_3/^{1}m^{2_{\frac{2}{3}\pi}}m^{2_{\frac{1}{3}\pi}}c^{m_{001}}1$). Remarkably, both results are in excellent agreement with experimental observations. As illustrated in Figs. 4(a) and 4(b), the energy differences between the lowest and the second-lowest magnetic configurations in MnTe and $Mn_3Sn$ both exceed 10 meV/mag. atom, substantially larger than that of $CoNb_3S_6$ (1.4 meV/mag. atom), indicating that their ground-state magnetic orders are more robust. This is qualitatively consistent with the above-room-temperature magnetic transition temperatures of MnTe and $Mn_3Sn$ recorded in MAGNDATA, which are much higher than the 28 K of $CoNb_3S_6$. Notably, in both MnTe and $Mn_3Sn$, the second-lowest-energy configurations are coplanar helical antiferromagnets along the $c$ axis with propagation vector (0, 0, 0.25), manifesting electronic structures with odd-parity spin textures. Importantly, similar modulated helical antiferromagnetic structures have also been experimentally observed in recent studies of $Mn_3Sn$, depending on the synthesis conditions [61–63]. Therefore, MnTe also hosts the potential for the switch between ground-state even-parity and odd-parity spin splitting patterns [64].

Following the identification of the ground magnetic geometries for MnTe and $Mn_3Sn$, we further use the SSG-MSG subgroup chains to generate magnetic structures with explicit spin orientations. The corresponding results from relativistic DFT calculations are shown in Figs. 4(c) and 4(d). In $Mn_3Sn$, the lowest-energy magnetic structure is found to possess *Cmc*'*m*' symmetry, while another nearly degenerate orientation with *Cm*'*cm*' symmetry lies only 0.1 μeV/mag. atom higher. Such an ultralow magnetic anisotropy energy quantitatively aligns with both prior theoretical analyses and recent experimental insights [65]. Both configurations exhibit in-plane weak ferromagnetic moments, consistent with the experimentally observed small and soft ferromagnetic component [66], and can be regarded as a manifestation of SOM. By contrast, in MnTe the calculated ground state possesses *Cmcm* symmetry, with a competing *Cm*'*c*'*m* orientation only 0.5 μeV/mag. atom above in energy, consistent with the experimental identification of MnTe as an easy-plane antiferromagnet [67]. Interestingly, the two MnTe orientations display distinct transport responses: *Cmcm* configuration forbids the anomalous Hall conductivity, whereas *Cm*'*c*'*m* permits it, indicating that the magnetic transport properties of MnTe can be tuned by external fields and exploited for antiferromagnetic memory devices [67–69].

Overall, the three representative examples demonstrate that the symmetry-guided workflow can efficiently determine magnetic ground states and identify low-energy metastable states with distinct physical properties. For $CoNb_3S_6$, MnTe and $Mn_3Sn$, the screening requires only 27, 33, and 48 SOC-free DFT calculations, supplemented by just 4, 3, and 4 SOC-included ones, respectively. Beyond reproducing the experimentally reported ground states, these calculations reveal energetically favorable metastable states with spin-split, topological, or transport responses, paving possible routes for external-field switching between different unconventional magnetic phases [70]. These results highlight that the present framework provides an efficient and reliable route for predicting magnetic configurations prior to costly experimental characterization.

## IV. Benchmark

After demonstrating the workflow on three representative unconventional magnets, we now perform large-scale blind tests to systematically evaluate the reliability and generality of the proposed symmetry-guided framework. The MAGNDATA database provides a collection of experimentally determined magnetic structures and therefore serves as an ideal benchmark platform for this verification [10]. In Sec. IV A, the validity of the symmetry-guided generation method is verified by examining whether the generated symmetry-allowed magnetic structures successfully encompass the experimentally reported structures. In Sec. IV B, we perform first-principles calculations to further assess whether the experimentally observed magnetic states are among the relatively low-energy configurations. We also analyze the calculated metastable phases and their symmetry-allowed emergent physical properties.

### A. Benchmark of symmetry-guided generation

We first benchmark the symmetry-guided generation method against experimentally reported magnetic structures in the MAGNDATA database. Here we start from 1807 stoichiometric magnetic structures as stated in Sec. IIC. Among them, magnetic structures failing to satisfy the pre-defined generation constraints described in Sec. IIC are further filtered out, including cases requiring magnetic cell expansions larger than four, higher-level effective SG ($G_0$) subgroups, or complex magnetic-order-induced Wyckoff-position splitting, yielding 1595 remaining magnetic structures. In addition, we exclude a minor fraction of only 0.8% (12) cases exhibiting partial zero moments within the same crystallographic orbit. Consequently, 1583 magnetic structures, which correspond to 1178 distinct crystal structures, form the benchmark dataset. Since our method generates magnetic structures solely from its crystallographic inputs, all subsequent statistics are evaluated based on the 1178 crystal structures.

For each crystal structure with the specified magnetic Wyckoff positions, our method exhaustively generates SSG-allowed magnetic geometries without requiring a priori magnetic inputs, yielding an average of 60 inequivalent magnetic geometries per crystal structure. To compare a generated geometry with the experimentally reported magnetic structure in MAGNDATA, we use the pairwise relative angles between magnetic moments. For a set of unit spin directions $\{\boldsymbol{e}_i\}$, the relative-angle matrix A is defined by $A_{ij} = \arccos(\boldsymbol{e}_i \cdot \boldsymbol{e}_j)$ with $A_{ii} = 0$. The geometry deviation angle is then evaluated as $\Delta\theta = \max(|A_{gen} - A_{exp}|)/2$, where $A_{gen}$ and $A_{exp}$ are the relative-angle matrices of the generated and experimental magnetic structures, respectively. This metric compares magnetic geometries at the level of relative spin angles and is independent of the global spin orientation.

Since experimental structures naturally include SOC effects, an exact match to the ideal SSG-generated geometry is neither expected nor required. For each experimental magnetic structure, we identify the SSG-generated magnetic geometry with the smallest geometry deviation angle. We then cross-examine the SSG-MSG subgroup chain of this matched SSG to check whether its MSG descendants include the experimentally reported MSG. A magnetic structure is classified as successfully recovered when the geometry deviation angle satisfies $\Delta\theta \leq 10^{\circ}$ and the experimental MSG is contained within the symmetry-breaking pathways of the matched SSG. Under these conditions, the small deviation can be physically interpreted as weak SOC-induced canting, e.g., driven by Dzyaloshinskii–Moriya interaction [71,72], on top of the exchange-dominant SSG-generated magnetic geometry.

Using this criterion, the experimentally reported magnetic structures are successfully recovered for 922 out of the 1178 crystal structures at the SSG level, achieving a 78% recovery rate. Figure 5(a) summarizes the distribution of the geometry deviation angles for the recovered structures, where the bars are further decomposed according to the matched MSG/OSSG type. Specifically, these structures are classified into easy-axis, easy-plane, and arbitrary-axis categories. Notably, 76% (699/922) of the recovered structures belong to the easy-axis category, where the matched OSSG completely locks the local magnetic moments along specific crystallographic axes. Conversely, the easy-plane and arbitrary-axis categories represent cases where the matched OSSG allows the orientation to vary continuously within a two-dimensional plane or a three-dimensional space to match the experimental MSG at general positions. During this continuous rotation, the corresponding MSG can either alter, such as by exhibiting enhanced symmetry along specific high-symmetry directions, or remain invariant. We identify 161 materials within these two categories that maintain a strictly invariant MSG throughout the continuous rotation. From an energetic perspective, these structures remain quasi-degenerate on the SOC-included energy landscape. Along these directions, the energy variations induced by SOC are typically negligible, causing the continuous spin rotation to exhibit a quasi-Goldstone mode. This behavior is most prevalent in systems featuring only a single axial operation, particularly under monoclinic crystal structures. For instance, in the collinear antiferromagnet $Sr_2ScOsO_6$ (#0.917 in MAGNDATA), the MSG remains unchanged when the Néel vector rotates within the plane perpendicular to the crystallographic principal axis. In such cases, distinguishing these nearly degenerate directions is unnecessary, which often default to an arbitrary high-symmetry direction in experiments [73]. Consequently, treating these cases as successful at the OSSG level is physically justified, raising the success rate to 93% (860/922).

The concentration of most recovered structures in the small-deviation-angle region confirms that the SSG-generated geometries successfully capture the dominant exchange-driven magnetic arrangements observed experimentally. Figure 5(a) illustrates the typical weak-ferromagnetic canting induced by Dzyaloshinskii-Moriya interaction [71,72]. Beyond such antisymmetric exchange interaction, this deviation can also originate from single-ion anisotropy, as seen in $CsMnI_3$ that exhibits a canting angle of 9.9º relative to the ideal 120º coplanar magnetic structure [Figure 5(b)] [74]. While we adopt the $\Delta\theta \leq 10^\circ$ threshold for our generation benchmark, SOC-driven deviation can certainly exceed this limit. For example, NdSi exhibits an antiferromagnetic canting angle of 20º relative to the ideal collinear ferromagnetic arrangement, arising from symmetric exchange anisotropy [75]. These examples illustrate that SOC-induced canting can naturally appear in experimentally refined magnetic structures, while the dominant magnetic geometry, governed primarily by isotropic exchange interactions, remains correctly captured by the SSG-generated state. Such secondary magnetic components can be partially captured in subsequent SOC-included DFT calculations, as demonstrated for $CoNb_3S_6$ in Sec. IIIA, where SOC yields a weak moment allowed by the MSG P32'1. For the remaining 22% (256/1178) of materials where the framework fails to recover the exact experimental structures, a special category consists of 2 structures that exhibit small canting angles below 10º but fail recovery because their experimentally reported MSG violates the SSG-MSG subgroup chain relation, as exemplified by $SrMnSb_2$ (#0.768 in MAGNDATA), which warrants further experimental and theoretical re-examination. The remaining majority, such as $Mn_3RhGe$ (#0.1005 in MAGNDATA), involve low-symmetry spin Wyckoff positions with multiple degrees of freedom of local magnetic moments, and are left for future study.

Overall, the benchmark against MAGNDATA demonstrates that the symmetry-guided generation method can effectively reproduce experimentally reported magnetic structures for the majority of the tested crystal structures, without any prior knowledge of their magnetic ordering. This validates the reliability of the OSSG-based generation scheme as an efficient approach for constructing candidate magnetic structures before first-principles energy ranking.

### B. First-principles total-energy benchmark and emergent metastable phases

To further assess the reliability of the first-principles energy-ranking procedure, we select a subset of 305 crystal structures from the 922 successfully recovered structures in the generation benchmark. These structures contain no *f*-electron elements, have no more than 20 atoms in the primitive cell, and involve only one crystallographic Wyckoff orbit of the magnetic atoms in the parent space group. This filtering primarily aims to accelerate the large-scale benchmark and avoid the difficulties of treating localized *f* electrons by standard DFT. For each selected crystal structure, we perform nonrelativistic first-principles calculations for the SSG-generated magnetic geometries and examine whether the experimentally reported magnetic structure is recovered among the computed low-energy configurations. The details of high-throughput calculations are provided in Appendix E. Given that magnetic energy differences are typically small and can be sensitive to computational parameters and structural details, the experimental magnetic structure is considered correctly predicted if it lies within 5 meV/mag. atom of the computed lowest-energy configuration.

The distribution of the calculated energy difference between the experimental magnetic state and the computed lowest-energy configuration is shown in Fig. 5(c); the correctly predicted magnetic structures and the rest of the calculated ones are listed in Tables I and II, respectively (see Appendix F). This criterion is satisfied for 82% of the tested structures, indicating that these magnetic structures are primarily governed by SOC-free exchange interactions. Representative cases of failures in the weak-SOC 3*d* systems include the van der Waals layered materials $FeCl_2$ and $FeBr_2$, for which the calculated energy differences are more than 100 meV/mag. atom above the computed ground states in the nonrelativistic calculations. This large energy discrepancy is closely associated with magnetic-order-dependent changes in the electronic structure, such as a metal-to-semiconductor transition. The experimentally observed magnetic structures of $FeCl_2$ and $FeBr_2$, characterized by intralayer ferromagnetic and interlayer antiferromagnetic order, remain metallic in the absence of SOC. When SOC is included, the orbital occupation is modified, resulting in a large-gap insulating state [76].

Beyond recovering experimentally reported magnetic configurations, the first-principles benchmark also reveals a substantial number of low-energy metastable magnetic phases. As summarized in Fig. 5(d), among the configurations lying within 5 meV/mag. atom of the computed lowest-energy state, we identify 106 altermagnetic configurations in 59 materials, 1176 spin-splitting noncollinear antiferromagnetic configurations in 174 materials, and 24 geometric-Hall antiferromagnetic configurations in 22 materials. Notably, among the spin-splitting noncollinear antiferromagnetic configurations, 1165 exhibit odd-parity spin textures, mainly arising from coplanar spiral-like magnetic geometries, while 5 exhibit even-parity spin textures. This demonstrates that the present framework not only predicts experimentally observed magnetic states, but also provides a systematic route to identify competing metastable phases with symmetry-allowed emergent properties.

Two representative examples are shown in Figs. 5(e) and 5(f). In $MnTe_2$, in addition to correctly predicting the experimentally observed spin-splitting noncoplanar antiferromagnetic geometry [77,78], our workflow identifies a low-energy collinear antiferromagnetic metastable state with SSG 61.14.1.1.L ($P^{-1}b^{-1}c^{1}a^{\infty m}1$). Interestingly, this magnetic structure has been reported in earlier studies [79], which can be explained by the fact the our calculated total energy is only 0.9 meV/mag. atom lower than that of the noncoplanar one. Furthermore, the collinear magnetic structure exhibits *d*-wave altermagnetic spin splitting [Fig. 5(e)]. The coexistence of the noncoplanar ground state and the collinear metastable state suggests that $MnTe_2$ lies near a phase boundary between competing exchange interactions. This is consistent with the previous work on how higher-order spin interactions play a key role in stabilizing noncollinear states against collinear states governed by Heisenberg model [79].

In $KRuO_4$, our framework correctly predicts the experimental collinear antiferromagnetic ground state [81] and further reveals a noncoplanar metastable state only 2.18 meV/mag. atom above the ground state. This corresponding SSG is 82.3.4.4 ($I^{2_{1-10}}\bar{4}|(2_{001}, 2_{001}, 1;\ 2_{100})$. Although the combined $PT$ symmetry is absent, where $P$ denotes spatial inversion, the band structure remains spin degenerate throughout the Brillouin zone. This degeneracy originates from two mutually perpendicular $U(\pi)\tau$ symmetries in the magnetic configuration, where $U(\pi)$ denotes a twofold spin rotation [47]. Symmetry analysis further shows that a finite orbital magnetization along the *z* direction is allowed, leading to a nonzero anomalous Hall conductivity $\sigma_{xy}$ without SOC. As shown in Fig. 5(d), metastable states exhibiting such geometric Hall effects are relatively rare. Finally, the identified metastable magnetic phases with emergent properties will be available to public through the FINDSPINGROUP website [48].

### V. Discussion

Our symmetry-guided framework represents a broader application of state-of-the-art spin group theory. Previous applications have primarily focused on known magnetic materials, where SSGs have been used to reveal novel physical properties in the absence of SOC [39]—such as spin splitting (as in altermagnets [6]), quantum geometric effects [82], and topological magnons [5]. However, here we address the fundamental inverse question, i.e., whether these decoupled spin and spatial symmetries can actively guide the discovery of new magnetic structures [12]. Compared with existing approaches that depend on empirical spin models or predefined magnetic propagation vectors, our framework requires only the crystal structure and magnetic ions to reliably screen the symmetry-adapted magnetic configuration space. Furthermore, this work provides an answer by demonstrating how spin-group symmetry can be leveraged to establish previously unreported magnetic structures, thereby also facilitating neutron-diffraction refinement by substantially reducing the number of independent fitting parameters compared with conventional MSG-based descriptions.

Given the predicted magnetic structures, our framework also provides insight on the underlying magnetic interactions in different systems. This capability is illustrated by MnTe and $Mn_3Sn$, which share the same nonmagnetic SG and magnetic atoms. However, by surveying the ten lowest-energy magnetic configurations, we find that in MnTe, most configurations favor interlayer antiferromagnetic coupling, whereas in $Mn_3Sn$, the majority favor intralayer antiferromagnetic coupling, leading to frustrated triangular spin arrangements. This divergence arises from the distinct exchange network: in MnTe, the dominant Mn–Mn exchange pathways are oriented primarily along the out-of-plane direction, while in $Mn_3Sn$ they are confined within the kagome planes. Such contrasting behavior not only accounts for the different spin geometries of these compounds but also underscores the critical role of lattice geometry and orbital overlap in shaping their magnetic energy landscapes. Additionally, our framework successfully captures low-energy competing states arising from temperature-dependent phase transitions within a single parent phase. These calculated configurations represent tunable metastable phases, offering viable pathways for manipulation via external fields [70].

The present framework also has limitations, which can be classified into two main categories. The first category concerns the coverage of the symmetry-guided generation step. First, while the current generation protocol targets commensurate magnetic structures within a fourfold cell expansion, SSG symmetry is fundamentally capable of guiding the exploration of incommensurate magnetic orders. Historically, the simplification of spin-spiral calculations leveraged SSG symmetry to bypass massive supercells [24]; therefore, utilizing SSG constraints to explore incommensurate magnetism represents a highly promising avenue for our future implementation. In practice, the low-energy configurations obtained within fourfold cell expansions provide useful physical guidance: for example, competing noncollinear states with different periodicities may indicate a physical tendency toward spiral order. These configurations can serve as training data for subsequent global searches using spin-spiral calculations or effective spin models. Second, when magnetic sites locate at spin Wyckoff positions with lower site symmetry, the magnetic moment can have multiple degrees of freedom. In such cases, symmetry alone cannot uniquely determine the relative spin directions, and handling these continuous parameters requires exhaustive geometric sampling that increases the complexity of structural generation. To circumvent this, these symmetry-determined structures can serve as anchors to parameterize spin Hamiltonians or train machine learning models, which can then optimize the continuous parameters. Third, our framework assumes that the emergence of magnetism is driven by the magnetic order and the subsequent SOC effects. It does not account for complex phase transitions driven by multi-order-parameter competition, such as strong magnetoelastic coupling, charge disproportionation, or do not obey the group-subgroup relations. Capturing such cases remains a formidable challenge, and thus lies beyond the scope of the current framework.

The second category addresses numerical and physical approximations from the perspective of DFT calculations. First, because the predictive reliability depends heavily on DFT total-energy comparisons, its accuracy remains constrained by computational parameters. For example, when several competing magnetic states are separated by only a few meV per magnetic atom, the energy ranking may depend sensitively on computational parameters and structural details. To mitigate this resolution limit, we adopt a conservative 5 meV per magnetic atom tolerance window to define plausible ground-state candidates rather than enforcing a strict absolute energy ranking. Finally, our current framework takes reported magnetic crystal structures as inputs, rather than directly from non-magnetic parent crystal structures. For workflows screening from non-magnetic crystal structures, while magnetic-order-induced lattice deformations are typically subtle, accounting for these deformations becomes critical when several competing low-energy states are nearly degenerate.

In summary, we have developed a symmetry-guided framework for predicting magnetic structures without requiring predefined spin Hamiltonians or magnetic propagation vectors. The extensive validation by the MAGNDATA database demonstrates the general capability of our framework to efficiently navigate complex magnetic configuration spaces without relying on empirical parameters. In the future, the systematic magnetic configurations generated here naturally provide an ideal foundation for fitting effective spin Hamiltonians and training machine-learning potentials. Ultimately, this framework is expected to establish a systematic foundation to explore the microscopic origins of unconventional magnetism, thereby paving the way for the targeted design of novel magnetic materials across diverse systems.

**Acknowledgments**

We thank Juan Manuel Perez-Mato and Xiangang Wan for helpful discussions. This work was supported by National Key R&D Program of China under Grant No. 2025YFA1411300, National Natural Science Foundation of China under Grants No. 12525410, No. 12274194, No. 12574275, and No. 12534003, and Guangdong Provincial Quantum Science Strategic Initiative under Grant No. GDZX2401002, Guangdong Provincial Key Laboratory for Computational Science and Material Design under Grant No. 2019B030301001, Shenzhen Science and Technology Program (Grant No. RCJC20221008092722009, No. 20231117091158001 and No. QNXMC20250701094702004), the Innovative Team of General Higher Educational Institutes in Guangdong Province (Grant No. 2020KCXTD001), China Postdoctoral Science Foundation under Grant No. 2025M783394, the Open Fund of the State Key Laboratory of Spintronics Devices and Technologies (Grant No. SPL-2407) and Center for Computational Science and Engineering of Southern University of Science and Technology.

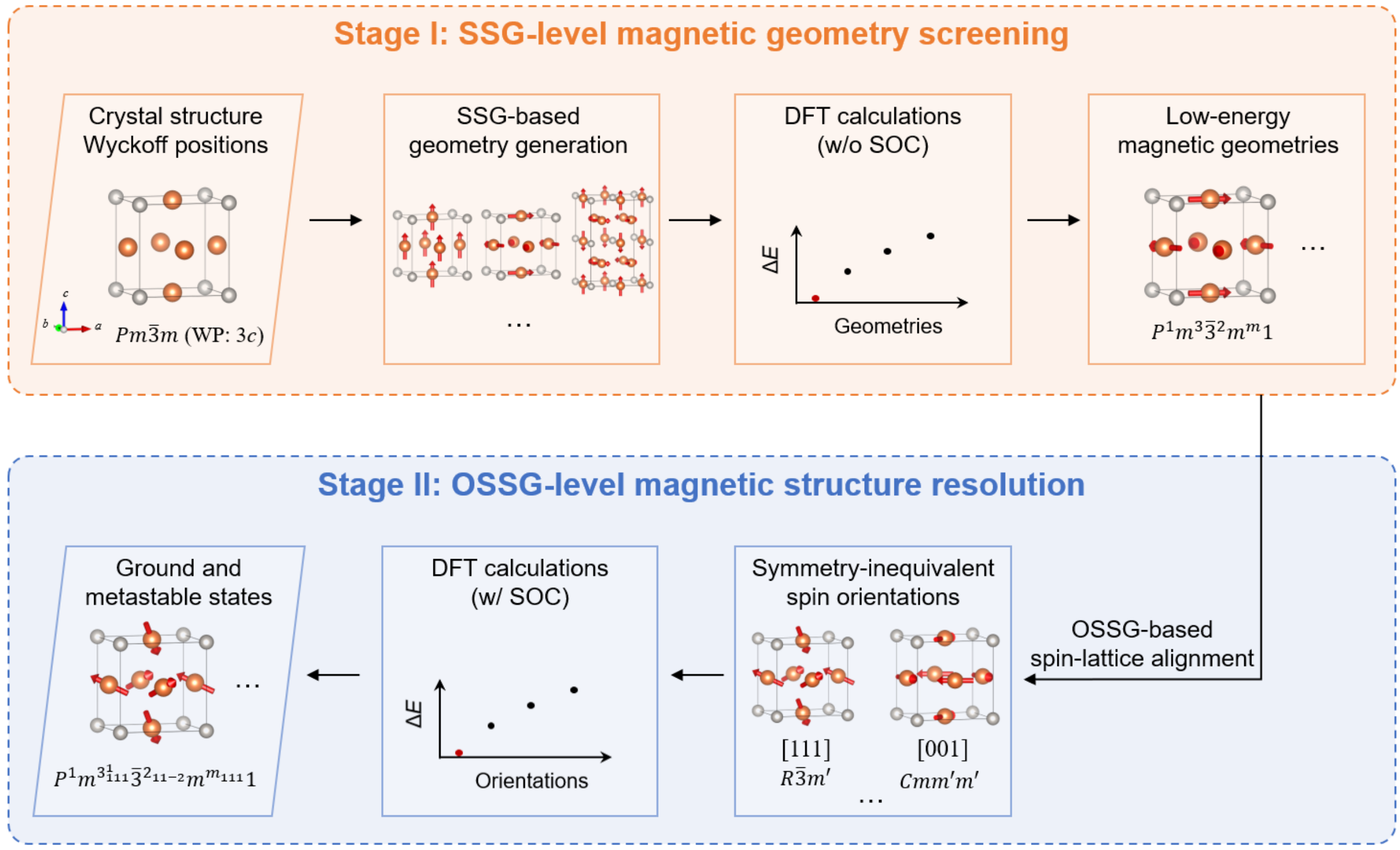

**FIG. 1. Workflow for predicting magnetic ground states.** In the SSG-level magnetic geometry screening (stage I), candidate magnetic geometries are systematically generated via SSG-based geometry enumerations and screened by DFT calculations without SOC to capture isotropic exchange interactions. In the OSSG-level magnetic structure resolution (stage II), the OSSG framework is introduced to reduce the continuous orientation space into a finite set of symmetry-inequivalent spin orientations, thereby incorporating SOC-induced magnetic anisotropy. In the illustrative example, for a parent crystal structure with space group $Pm\bar{3}m$ and magnetic atoms occupying the 3$c$ Wyckoff position (WP), the workflow successfully determines the experimental ground magnetic structure characterized by the OSSG $P^{1}m^{3^{1}_{111}}\bar{3}^{2_{11-2}}m^{m_{111}}1$.

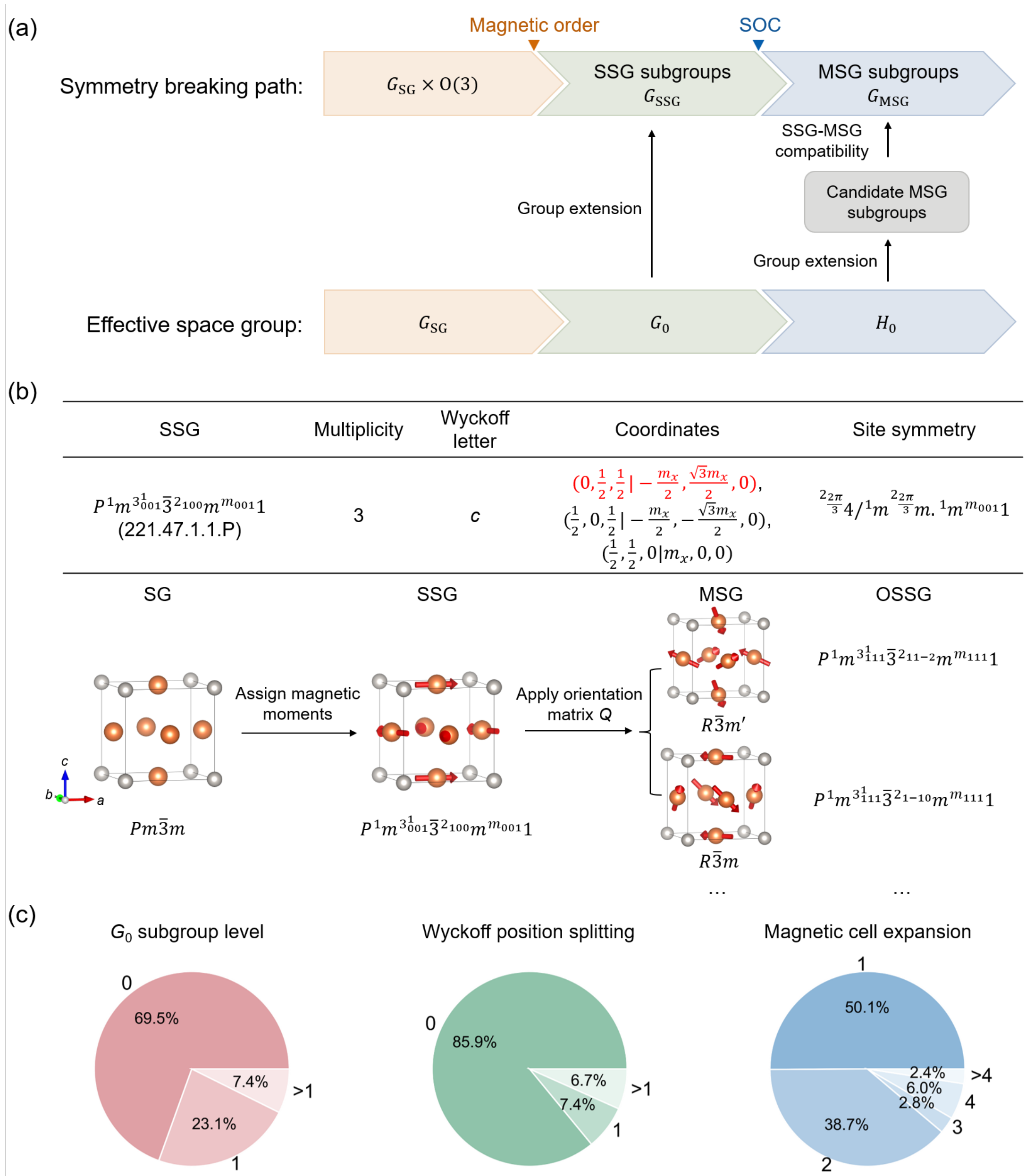

**FIG. 2. Construction of SG-SSG-MSG subgroup chains and symmetry-guided generation of magnetic structures.** (a) Subgroup-chain construction from the parent crystallographic space group $G_{\mathrm{SG}}$. SSG subgroups are generated from the effective space group $G_0 \subseteq G_{\mathrm{SG}}$, while candidate MSG subgroups are generated from $H_0 \subseteq G_0$ and selected by SSG-MSG compatibility through the orientation matrix $Q$. (b) Example of magnetic-structure generation from a spin Wyckoff position of an SSG. The SSG $P^{1}m^{3^{1}_{001}}\bar{3}^{2_{100}}m^{m_{001}}1$ is also labeled as 221.47.1.1.P using the Chen-Liu symbol ($G_0.L_0.i_k.$index.L/P) from ref. [48], where index represent the mapping sequence in the database. The additional index L or P represent a collinear or coplanar spin-only group, respectively, and is omitted for non-coplanar configurations. The 3$c$ spin Wyckoff position shown in the table displays the symmetry-constrained site coordinates and magnetic moments, as well as the spin site-symmetry group of the representative site (highlighted in red). Assigning these moments to the magnetic sites generates a magnetic geometry, and applying $Q$ further produces MSG-allowed spin orientations. (c) MAGNDATA statistics motivating the generation constraints on $G_0$ subgroup level, magnetic order-induced Wyckoff-position splitting of magnetic ions, and magnetic cell expansion.

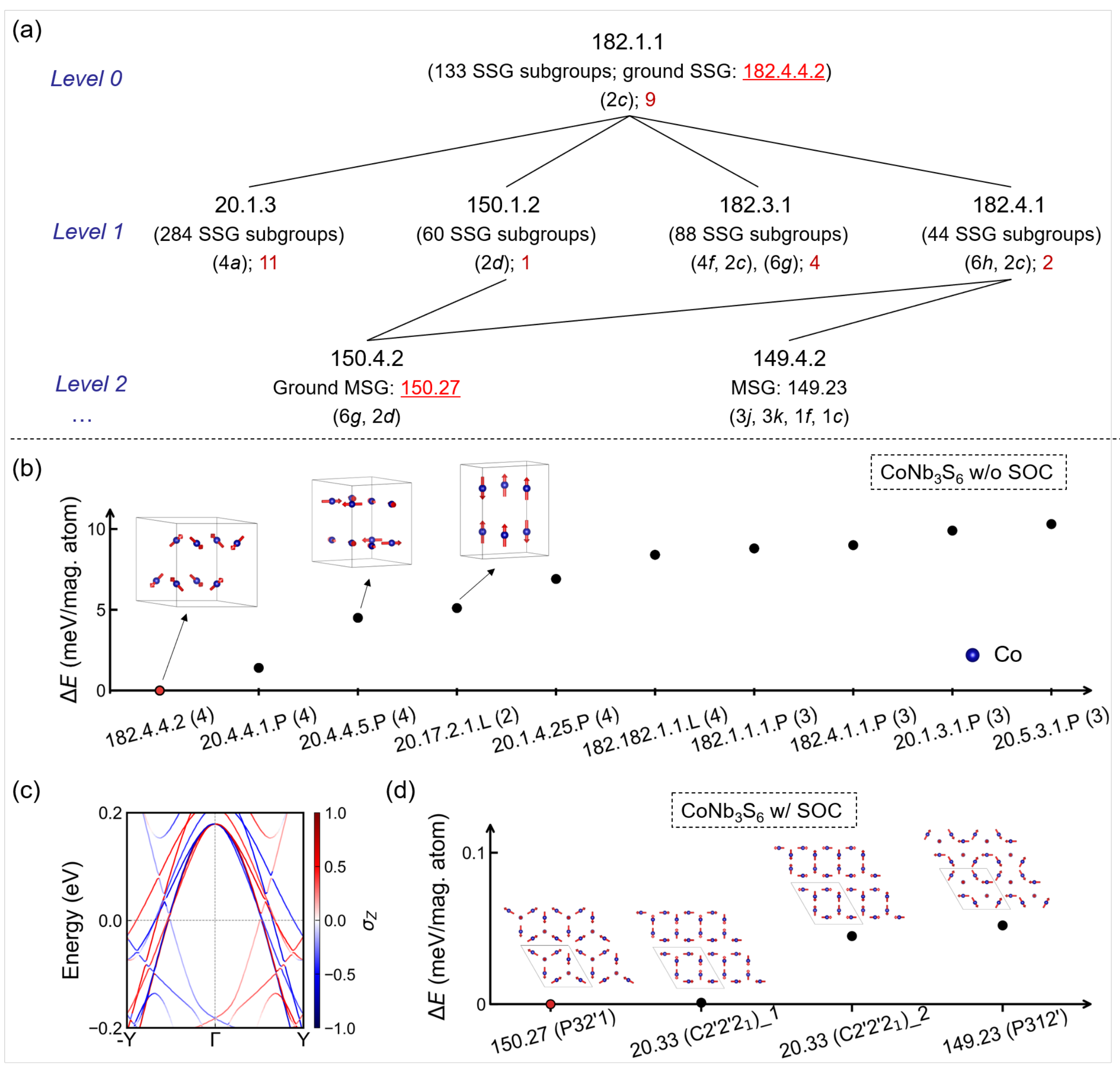

**FIG. 3. Magnetic ground state determination for $CoNb_3S_6$.** (a) A selected portion of SG-SSG-MSG subgroup chains generated from the parent crystallographic space group No. 182. Each effective space subgroup is labeled by its international number, subgroup $k$ index and subgroup $t$ index. The numbers of SSG subgroups derived from each effective space subgroup under the magnetic cell expansion cutoff of four are listed below the corresponding nodes. The magnetic-order-induced Wyckoff splitting of magnetic atoms and the resulting inequivalent magnetic geometries generated under the corresponding SSG subgroups are indicated. For the SSG subgroup 182.4.4.2, two compatible MSG subgroups are marked, with MSGs labeled by their BNS numbers. (b) Relative energies of the ten lowest-energy magnetic geometries, with selected representative spin structures shown in the insets. SOC is not included in these calculations. Only magnetic atoms are displayed. Each magnetic geometry is denoted by $G_{\text{SSG}}\ (n)$, where SSG is represented by the Chen-Liu nomenclature and $n$ corresponds to the magnetic cell expansion factor. (c) $S_z$-projected band structures of the 20.4.4.5.P (4)

magnetic configuration in the absence of SOC. (d) Calculated magnetic orientations, arranged from left to right in order of increasing total energy. Each magnetic structure is labeled by the BNS number and the international symbol of the corresponding MSG. SOC is included in the calculations. Relative energies are given in meV per magnetic atom.

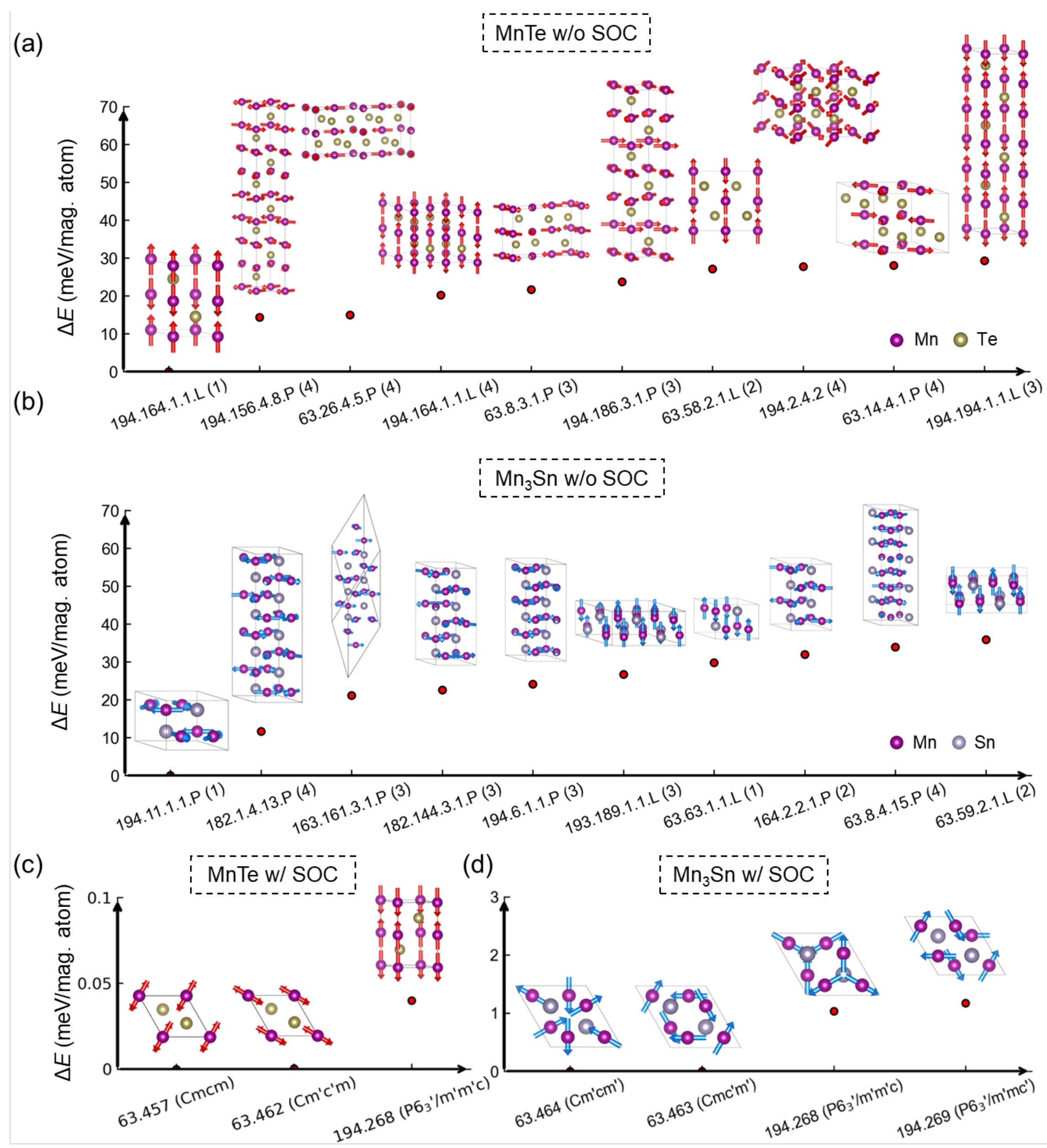

**FIG. 4. Magnetic ground state determination for MnTe and $Mn_3Sn$.** (a, b) Calculated relative energies of the ten lowest-energy magnetic geometries for MnTe and $Mn_3Sn$, respectively. Each magnetic geometry is denoted by $G_{\mathrm{SSG}}\,(n)$, as defined in Fig. 3. (c, d) Calculated magnetic orientations for MnTe and $Mn_3Sn$, respectively, arranged from left to right in order of increasing total energy. Each magnetic orientation is labeled by the BNS number and the international symbol of the corresponding MSG. Relative energies are given in meV per magnetic atom.

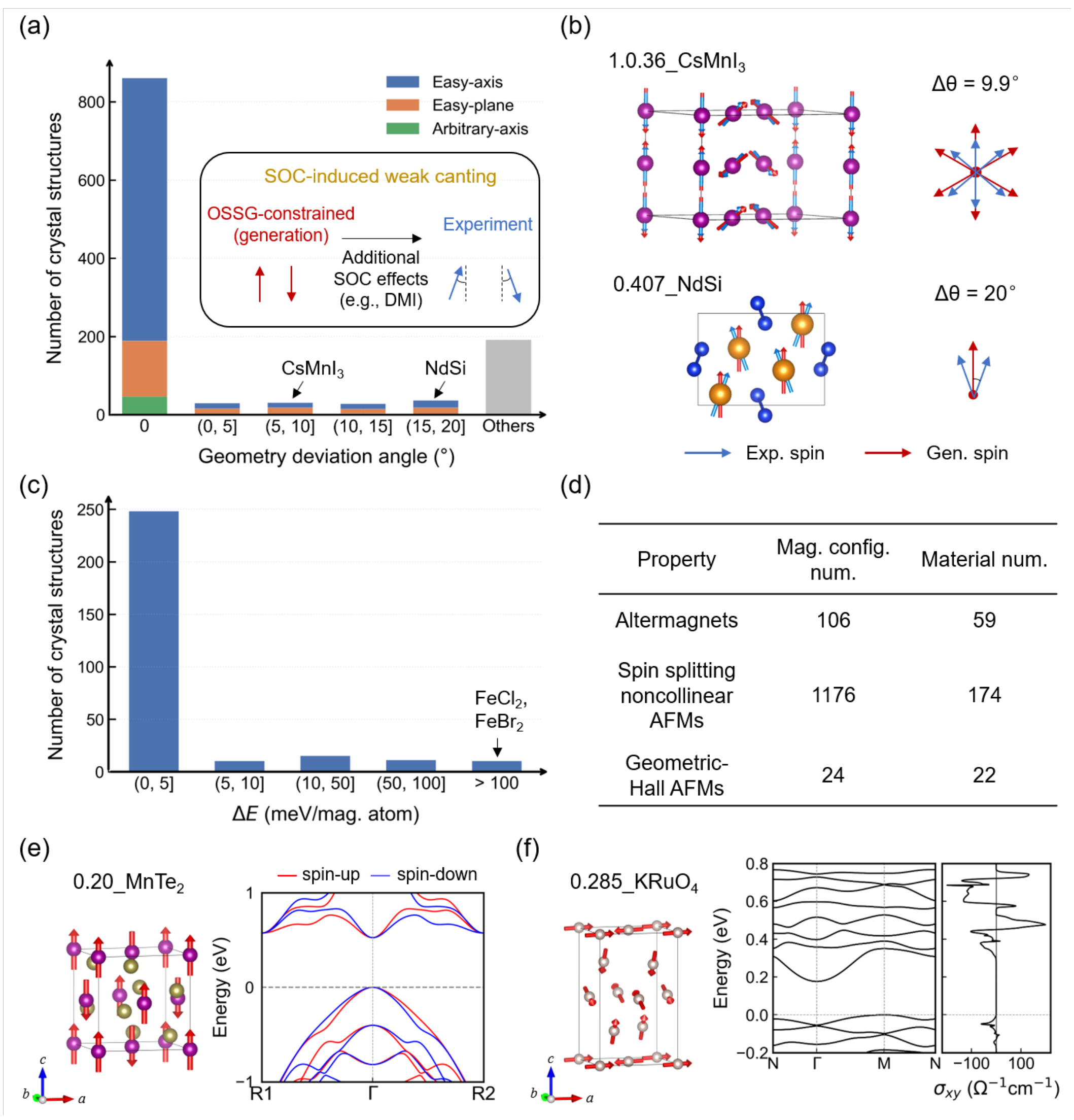

| Property | Mag. config. num. | Material num. |
|---|---|---|
| Altermagnets | 106 | 59 |
| Spin splitting noncollinear AFMs | 1176 | 174 |
| Geometric-Hall AFMs | 24 | 22 |

**FIG. 5. Benchmark against the magnetic structures in MAGNDATA.** (a) Distribution of the geometry deviation angles between SSG-generated and experimentally reported magnetic structures, decomposed by MSG/OSSG-constrained orientation type. The inset illustrates SOC-induced weak canting from the OSSG-constrained generated structure to the experimentally refined structure. (b) Representative examples, $CsMnI_3$ and NdSi, with blue and red arrows denoting experimental and generated magnetic moments, respectively. (c) Distribution of the calculated energy difference between the experimentally observed magnetic state and the computed lowest-energy configuration in the first-principles benchmark. (d) Summary of low-energy metastable configurations within 5 meV/mag. atom, classified by symmetry-allowed emergent properties. (e) Metastable altermagnetic state in $MnTe_2$. (f) Noncoplanar metastable state in $KRuO_4$ with nonzero geometric Hall conductivity in the absence of SOC.

## APPENDIX A: Comparison of magnetic space group and oriented spin space group sampling

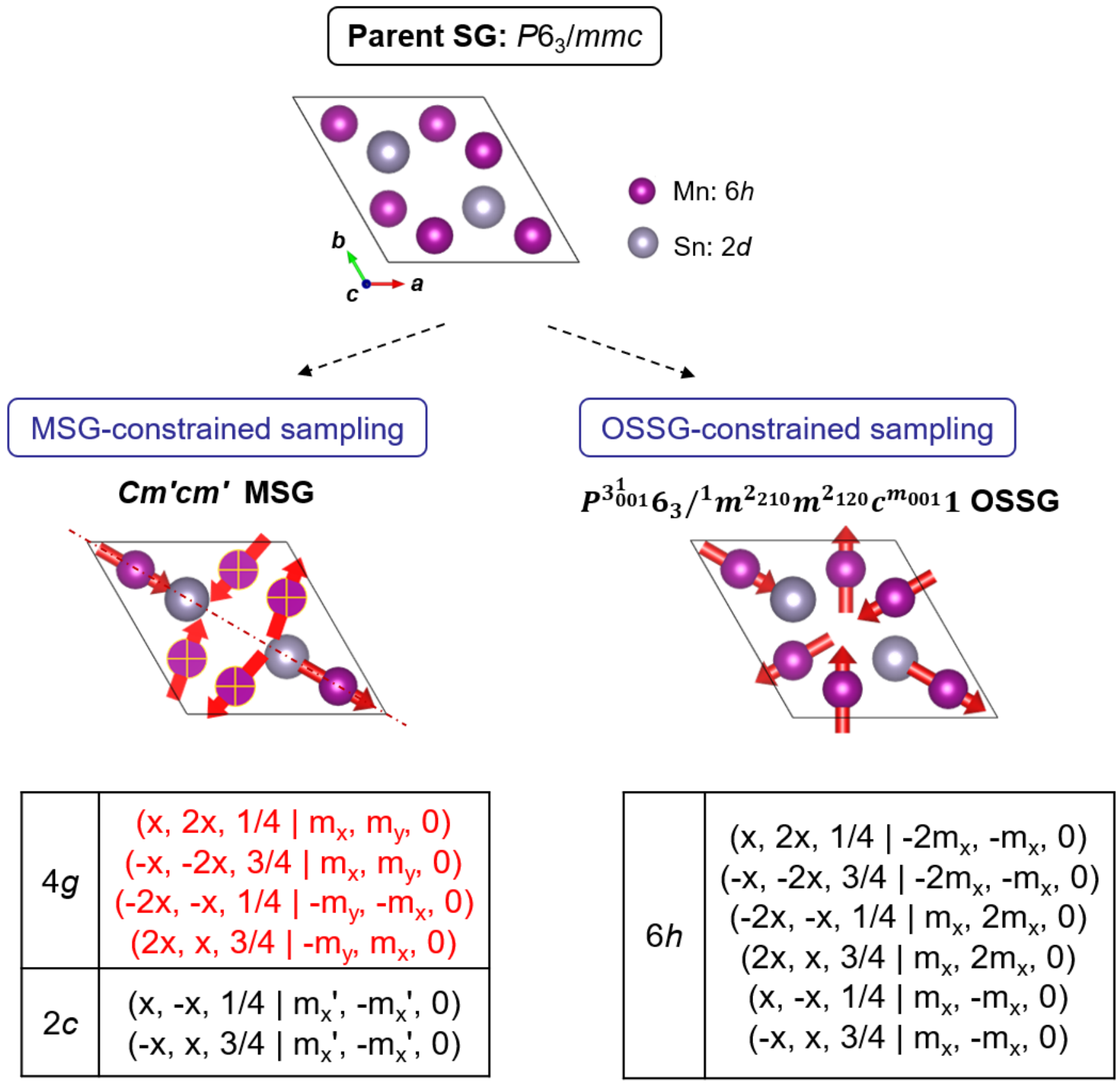

| | |
|---|---|
| 4*g* | (x, 2x, 1/4 \| $m_x$, $m_y$, 0)<br>(-x, -2x, 3/4 \| $m_x$, $m_y$, 0)<br>(-2x, -x, 1/4 \| $-m_y$, $-m_x$, 0)<br>(2x, x, 3/4 \| $-m_y$, $m_x$, 0) |
| 2*c* | (x, -x, 1/4 \| $m_x'$, $-m_x'$, 0)<br>(-x, x, 3/4 \| $m_x'$, $-m_x'$, 0) |
| 6*h* | (x, 2x, 1/4 \| $-2m_x$, $-m_x$, 0)<br>(-x, -2x, 3/4 \| $-2m_x$, $-m_x$, 0)<br>(-2x, -x, 1/4 \| $m_x$, $2m_x$, 0)<br>(2x, x, 3/4 \| $m_x$, $2m_x$, 0)<br>(x, -x, 1/4 \| $m_x$, $-m_x$, 0)<br>(-x, x, 3/4 \| $m_x$, $-m_x$, 0) |

**FIG. 6. Comparison of magnetic structure sampling under MSG and OSSG constraints for $Mn_3Sn$.** Under the MSG constraint, the Mn 6*h* Wyckoff orbit of SG splits into MSG symmetry-inequivalent 4*g* and 2*c* orbits, which are visually distinguished as the crossed-marked purple spheres and the plain purple spheres, respectively. Consequently, the overall magnetic moments are parameterized by three independent parameters. Under the OSSG constraint, the Mn 6*h* orbit remains unsplit and only one moment-magnitude parameter is required, as the 120° triangular magnetic geometry is uniquely dictated by symmetry.

To elucidate the distinction between the sampling of magnetic configuration space constrained by conventional MSGs and OSSGs, we examine the prototypical antiferromagnet $Mn_3Sn$. $Mn_3Sn$ crystallizes in the SG $P6_3/mmc$, with magnetic Mn atoms occupying the 6*h* Wyckoff position, and hosts an in-plane frustrated triangular antiferromagnetic order. One experimentally reported magnetic structure has *Cm'cm'* MSG symmetry, which corresponds to the entry 0.200 within the MAGNDATA database. As shown in Fig. 6, under the MSG constraint, the parent Mn 6*h* orbit splits into two symmetry-inequivalent 4*g* and 2*c* orbits, which correspond to the crossed-marked purple spheres and the plain purple spheres, respectively. As a result, the overall magnetic moments are described by three independent parameters. Two of them correspond to the moment-magnitude degrees of freedom of the 4*g* and 2*c* orbits, while the remaining parameter controls the relative spin angle between the moments. Although the MSG constraint reduces the 18-dimensional magnetic configuration space to a 3-dimensional subspace, it does not fully determine the triangular magnetic geometry. Direct MSG-constrained sampling therefore still leaves a residual continuous magnetic-geometry degree of freedom. In DFT-based searching of magnetic structures, this artificial geometric flexibility frequently expands the optimization space, leading to poor convergence or entrapment in local energy minima.

The same magnetic structure can also be described by the OSSG $P^{3^1_{001}}6_3/^{1}m^{2_{210}}m^{2_{120}}c^{m_{001}}1$. In contrast to the MSG description, the OSSG framework preserves the parent 6*h* Wyckoff position of Mn atoms as a single unsplit orbit. The relative 120° triangular spin arrangement is fully fixed by OSSG symmetry, leaving only one free parameter corresponding to the overall moment magnitude. Therefore, the OSSG constraint reduces the 18-dimensional magnetic configuration space to a 1-dimensional space, fully reproducing the target magnetic structure without geometrical ambiguity.

This comparison illustrates the advantage of OSSG-based sampling over direct MSG-based sampling. The OSSG can be regarded as a parent symmetry framework of the MSG, since its generators include $\{3^1_{001}||6^1_{001}|00\frac{1}{2}\}$, $\{1||m_{001}|00\frac{1}{2}\}$, $\{2_{210}||m_{100}|0\}$, $\{2_{120}||m_{210}|00\frac{1}{2}\}$ and $\{m_{001}||1|0\}$. These operations yield fully coupled spin-lattice operations such as $\{1||-1|0\}$, $\{m_{110}||m_{110}|0\}$, $\{m_{001}||m_{001}|00\frac{1}{2}\}$, which subsequently generate the MSG. In addition, the OSSG contains spin-lattice decoupled operations, such as $\{3^1_{001}||6^1_{001}|00\frac{1}{2}\}$, which are absent in the MSG description but are essential for capturing the geometric symmetry of the 120° triangular magnetic structure. Thus, compared with direct MSG-constrained sampling, OSSG-constrained sampling provides a more symmetry-refined description of magnetic structures and enables a finer partitioning of the high-dimensional magnetic configuration space. Consequently, OSSG-constrained sampling serves as a more efficient partitioning tool that directly locks the exact magnetic geometry, thereby significantly accelerating the high-throughput screening and discovery of complex magnetic structures.

**APPENDIX B: Conjugate classes of SSG subgroups and MSG subgroups**

In the enumeration of SSG and MSG subgroups, it is important to distinguish between group types and subgroup conjugate classes. An SSG or MSG type is mathematically determined by its group structure, such as the multiplication relations, the normal-subgroup decomposition, and the associated homomorphic mappings or group extensions. To data, more than 100,000 SSGs and 1,421 MSGs have been classified, providing the possible symmetry landscapes for magnetic structures. Additionally, the non-magnetic groups corresponding to 230 SGs are also incorporated, both without and with SOC, to provide reference configurations for DFT screening.

However, for magnetic-structure generation within a specific crystal, identifying the group alone is insufficient. For example, in a crystal with parent SG $P1$, magnetic structures generated by doubling the magnetic cell along the crystallographic *a*, *b*, *c* or other direction are physically distinct, because these directions are not related by any symmetry operation of the parent $P1$ group. Nevertheless, upon an appropriate change of basis, these different magnetic structures can be mapped onto the same standardized SSG setting and are therefore classified as the same SSG type $P^{1}1^{-1}(1/2,0,0)$ (1.1.2.1). Here, (1/2, 0, 0) also labels the propagation vector in the standardized setting of this SSG. Thus, classifying SSG and MSG subgroups only by their group type, without distinguishing the underlying subgroup embeddings, would fail to capture symmetry-inequivalent magnetic structures.

In the present framework, we resolve this ambiguity by distinguishing different SSG and MSG subgroup embeddings within the parent SG. With the standardized SSG and MSG group structures fixed, the subgroup embeddings are mainly specified by the transformation matrices relating the parent SG $G_{\mathrm{SG}}$ to effective SG subgroup $G_0$ or $H_0$. Different choices of these transformation matrices may generate distinct SSG or MSG subgroup classes, even when they correspond to the same SSG or MSG type. For the $P1$ example discussed above, the SSG type $P^{1}1^{-1}(1/2,0,0)$ (1.1.2.1) yields seven distinct classes of conjugate SSG subgroups, corresponding to the magnetic propagation vectors (1/2, 0, 0), (0, 1/2, 0), (0, 0, 1/2), (1/2, 1/2, 0), (1/2, 0, 1/2), (0, 1/2, 1/2), and (1/2, 1/2, 1/2), respectively.

Specifically, we use the following conjugate criteria to classify conjugate classes of SSG and MSG subgroups. Two SSG subgroups $G_{\mathrm{SSG1}}$ and $G_{\mathrm{SSG2}}$ are conjugate if there exists an operation $\{U||R|\boldsymbol{\tau}\} \in G_{\mathrm{SG}} \times \mathrm{O}(3)$ such that

$$\{U||R|\boldsymbol{\tau}\}^{-1} G_{\mathrm{SSG1}} \{U||R|\boldsymbol{\tau}\} = G_{\mathrm{SSG2}}. \tag{B1}$$

Similarly, two MSG subgroups $G_{\mathrm{MSG1}}$ and $G_{\mathrm{MSG2}}$ are conjugate if there exists an operation $\{R|\boldsymbol{\tau}\} \in G_{\mathrm{SG}}$ such that

$$\{R|\boldsymbol{\tau}\}^{-1} G_{\mathrm{MSG1}} \{R|\boldsymbol{\tau}\} = G_{\mathrm{MSG2}}. \tag{B2}$$

Conjugate SSG or MSG subgroups generate symmetry-equivalent magnetic structures and are therefore redundant. Physically, these conjugate subgroups correspond to degenerate magnetic domains with identical free energy, whereas non-conjugate classes represent fundamentally distinct magnetic phases. In our enumeration, one representative from each conjugate class of SSG and MSG subgroups is retained to exhaustively generate symmetry-inequivalent magnetic structures, which avoids redundant computational costs in subsequent high-throughput DFT screenings.

## APPENDIX C: SSG-MSG subgroup chains and associated OSSGs

In this appendix, we delineate the procedure for selecting compatible MSG subgroups for a given SSG subgroup and determining the corresponding OSSG orientation matrix $Q$. This step establishes the OSSG as a bridge connecting the SOC-free magnetic geometry dictated by an SSG to the SOC-constrained magnetic structure described by an MSG.

For a given SSG subgroup, candidate MSG subgroups are first generated from effective SG subgroups $H_0 \subseteq G_0$. To determine whether a candidate MSG is compatible with the SSG, we compare their symmetry operations with matched spatial parts. Specifically, for all spatially matched operation pairs ($\{U||R|\tau\}$, $\theta\{R|\tau\}$), we test whether the spin rotation $U$ in the SSG can be aligned, through a single global orientation matrix $Q$, with the spin transformation $\eta(\theta)\det(R)\,R$ imposed by the MSG operation, as required in Eq. (3). Before solving for $Q$, several necessary conditions are evaluated as a preliminary filter: namely, $U$ and $\eta(\theta)\det(R)\,R$ must have the same determinant and the same rotation order. Operation pairs that fail these conditions are incompatible, and the corresponding candidate MSG is discarded. If these conditions are satisfied, we further determine whether a $Q$ exists for all spatially matched operation pairs.

Rather than solving Eq. (3) directly, we evaluate this relation geometrically by comparing the rotation axes of the paired operations. Let $\hat{\boldsymbol{n}}_i^{(S)}$ and $\hat{\boldsymbol{n}}_i^{(M)}$ denote the rotation axes of $U_i$ and $R_i$, respectively. Since a rotation axis is intrinsically an unoriented line, Eq. (3) therefore requires that $Q$ maps each spin-space rotation axis onto the corresponding lattice-space rotation axis up to a sign,

$$Q\hat{\boldsymbol{n}}_i^{(S)} = \xi_i\hat{\boldsymbol{n}}_i^{(M)}, \qquad \xi_i = \pm 1. \tag{C1}$$

The relative angular relations among all matched rotation axes must also be preserved up to this sign ambiguity. In practice, this is checked by requiring

$$\left|\hat{\boldsymbol{n}}_i^{(S)} \cdot \hat{\boldsymbol{n}}_j^{(S)}\right| = \left|\hat{\boldsymbol{n}}_i^{(M)} \cdot \hat{\boldsymbol{n}}_j^{(M)}\right|, \tag{C2}$$

for all pairs of matched axes. Candidate MSG subgroups that fail this unoriented-axis consistency criterion are discarded.

The form of $Q$ depends on the number of independent matched rotation axes. If no matched rotation axis exists, for example when the effective SG of the MSG is *P*1 or *P*-1, $Q$ remains an arbitrary rotation matrix. For a single pair of matched axes, $Q$ can be written as a Rodrigues rotation that maps $\hat{\boldsymbol{n}}_i^{(S)}$ to $\hat{\boldsymbol{n}}_i^{(M)}$, followed by an arbitrary SO(2) rotation about $\hat{\boldsymbol{n}}_i^{(M)}$, which physically corresponds to a continuous degeneracy of the spin orientation. When two noncollinear matched axis pairs are available, $Q$ can be uniquely determined. The compatible MSG subgroups, together with the corresponding solutions of $Q$, constrain the allowed OSSG orientations and complete the SSG-MSG subgroup chains used for refined spin-orientation generation. The magnetic moments in the lattice frame are then obtained by applying the $Q$ to all SSG-generated moments,

$$\boldsymbol{M}_k = Q\boldsymbol{m}_k, \tag{C3}$$

where $\boldsymbol{m}_k$ is the magnetic moment generated in the standardized SSG spin frame, and $\boldsymbol{M}_k$ is the corresponding oriented magnetic moment in the lattice frame.

## APPENDIX D: Spin Wyckoff position of SSG

The spin Wyckoff positions of the SSG are derived by decomposing the SSG into the spin site-symmetry group of a representative site and its corresponding left cosets,

$$G_{\mathrm{SSG}} = \cup_{k=1}^{n}\{U^k||R^k|\boldsymbol{\tau}^k\}G_{s_0}, \tag{D1}$$

where $G_{s_0} = \left\{\{U^0||R^0|\boldsymbol{\tau}^0\} : R^0\boldsymbol{r}_0 + \boldsymbol{\tau}^0 = \boldsymbol{r}_0\right\}$ is the spin site-symmetry group that leaves the representative site $s_0$ invariant. Each coset representative $\{U^k||R^k|\boldsymbol{\tau}^k\}$ maps $s_0$ to a symmetry-related site $s_k$, acting simultaneously on the spatial coordinate and the magnetic moment,

$$\{R^k|\boldsymbol{\tau}^k\}\boldsymbol{r}_0 = \boldsymbol{r}_k, \qquad U^k\boldsymbol{m}_0 = \boldsymbol{m}_k. \tag{D2}$$

Here, $\boldsymbol{r}_0$ and $\boldsymbol{r}_k$ are the coordinates of sites $s_0$ and $s_k$, respectively, and $\boldsymbol{m}_0$ and $\boldsymbol{m}_k$ are their corresponding magnetic moments. The atomic position $\boldsymbol{r}_0$ and local magnetic moment $\boldsymbol{m}_0$ of $s_0$ are constrained by space operations $\{R^0|\boldsymbol{\tau}^0\}$ and spin operations $U^0$ in $G_{s_0}$, including spin-only operations, namely

$$\{R^0|\boldsymbol{\tau}^0\}\boldsymbol{r}_0 = \boldsymbol{r}_0, \qquad U^0\boldsymbol{m}_0 = \boldsymbol{m}_0. \tag{D3}$$

Thus, a spin Wyckoff position specifies not only the symmetry-allowed atomic coordinates, but also the corresponding symmetry-allowed magnetic moments. This construction enables the systematic generation of a standardized database of spin Wyckoff positions of SSG, which has been organized in our online database: https://database.findspingroup.com [48].

## APPENDIX E: Calculation details

First-principles calculations are performed using projector augmented-wave (PAW) method [83], as implemented in the Vienna *Ab initio* Simulation Package (VASP) [84]. Exchange-correlation effects are treated using the generalized gradient approximation (GGA) in the Perdew–Burke–Ernzerhof (PBE) functional form [85]. A plane-wave energy cutoff of 520 eV is used. Noncollinear calculations without SOC are performed to identify low-energy magnetic geometries, using a k-point density of 0.126 $Å^{-1}$ and an electronic self-consistency convergence criterion of $10^{-6}$ eV. SOC-included calculations are subsequently performed to determine the preferred magnetic orientations, using a denser k-point density of 0.063 $Å^{-1}$ and a stricter electronic convergence criterion of $10^{-8}$ eV. To account for strong electron correlation effects, the DFT+U method [86] is applied with $U_{\mathrm{eff}}$ = 1.0 eV for Co 3*d* electrons in $CoNb_3S_6$ [40] and $U_{\mathrm{eff}}$ = 3.03 eV for Mn 3*d* in MnTe [69]. No U correction is applied for $Mn_3Sn$. For the high-throughput benchmark calculations, the Hubbard *U* and Hund *J* parameters are adopted from the averaged values reported in the high-throughput linear-response study [87]. The crystallographic space groups of the input structures are identified using the spglib package [88,89]. The magnetic structure files (.mcif) are parsed using the pymatgen package [90]. Crystal structures and magnetic configurations are visualized using VESTA [91].

## APPENDIX F: High-throughput calculation results

Here, we provide the detailed high-throughput first-principles benchmark results discussed in Sec. IVB. Table I lists the materials for which the experimentally observed magnetic state is successfully recovered within 5 meV/mag. atom of the computed lowest-energy configuration. Table II lists the remaining cases for which the experimentally observed magnetic state lies outside this energy window.

TABLE I. Correctly predicted materials in the first-principles total-energy benchmark, where the calculated total energy of the experimental magnetic structure is less than 5 meV/mag. atom higher than that of the computed lowest-energy configuration. Each entry is designated by its standardized ID in the MAGNDATA repository followed by its chemical formula.

| Materials | | | | |
|---|---|---|---|---|
| 0.1_$LaMnO_3$ | 0.25_$NaOsO_3$ | 0.307_$ScCrO_3$ | 0.308_$InCrO_3$ | 0.309_$TlCrO_3$ |
| 0.323_$LaCrO_3$ | 0.432_$KMnF_3$ | 0.513_$YRuO_3$ | 0.586_$YCrO_3$ | 0.732_$SrRuO_3$ |
| 0.787_$YVO_3$ | 0.18_$BaMn_2As_2$ | 0.89_$BaMn_2Bi_2$ | 0.365_$BaCr_2As_2$ | 0.464_$BaMn_2P_2$ |
| 0.470_$BaMn_2Sb_2$ | 0.603_$CaMn_2Ge_2$ | 0.605_$BaMn_2Ge_2$ | 1.16_$BaFe_2As_2$ | 1.52_$CaFe_2As_2$ |
| 0.148_$La_2LiRuO_6$ | 0.301_$Sr_2CoTeO_6$ | 0.795_$Sr_2YRuO_6$ | 0.917_$Sr_2ScOsO_6$ | 0.934_$Sr_2NiTeO_6$ |
| 0.936_$Sr_2MnTeO_6$ | 1.166_$La_2LiOsO_6$ | 1.320_$Sr_2FeWO_6$ | 1.715_$Sr_2CoWO_6$ | 1.719_$Ca_2MnWO_6$ |
| 0.599_CaMnSi | 0.601_CaMnGe | 0.617_KMnSb | 0.618_KMnBi | 0.626_NaMnP |
| 0.629_NaMnAs | 0.631_NaMnSb | 0.634_NaMnBi | 0.222_CuMnAs | 1.539_KMnP |
| 1.541_RbMnP | 1.543_RbMnAs | 1.545_RbMnBi | 1.546_CsMnBi | 1.553_KMnAs |
| 0.58_$CoAl_2O_4$ | 0.462_$MnAl_2O_4$ | 1.138_$MgV_2O_4$ | 1.185_$GeCu_2O_4$ | 1.562_$GeNi_2O_4$ |
| 1.564_$GeCo_2O_4$ | 0.619_LaMnAsO | 0.667_LaMnSbO | 1.125_LaFeAsO | 1.146_LaCrAsO |
| 1.31_MnO | 1.619_MnS | 1.799_MnO | 1.6_NiO | 0.92_$CaMn_2Sb_2$ |
| 0.482_$SrMn_2As_2$ | 0.74_$Mn_3Cu_{0.5}Ge_{0.5}N$ | 0.177_$Mn_3GaN$ | 1.250_$KNiF_3$ | 1.809_$BiCoO_3$ |
| 0.273_$Mn_3ZnN$ | 0.142_$Fe_2TeO_6$ | 0.76_$Cr_2TeO_6$ | 0.75_$Cr_2WO_6$ | 0.966_$V_2WO_6$ |
| 2.22_$FeTa_2O_6$ | 1.0.41_$RbNiCl_3$ | 1.0.35_$CsMnBr_3$ | 1.0.36_$CsMnI_3$ | 1.0.9_$CsCoCl_3$ |
| 1.0.39_$BaMnO_3$ | 1.0.42_$CsNiCl_3$ | 1.527_$CsNiF_3$ | 0.1091_$La_2O_3Mn_2Se_2$ | 1.168_$Sr_2CuTeO_6$ |
| 1.433_$Ba_2YRuO_6$ | 1.538_$Ba_2MnTeO_6$ | 1.706_$Ba_2MnTeO_6$ | 1.707_$Ba_2MnWO_6$ | 0.699_$LiMn_6Sn_6$ |
| 0.778_$LaMnSi_2$ | 1.380_$Sr_2FeO_3Cl$ | 1.381_$Sr_2FeO_3Br$ | 1.382_$Ca_2FeO_3Cl$ | 1.383_$Ca_2FeO_3Br$ |

| 1.386_$Sr_2FeO_3F$ | 1.389_$Sr_2CoO_3Cl$ | 1.842_$Ca_2FeO_3Cl$ | 1.845_$Sr_2FeO_3Cl$ | 1.158_$YMn_3Al_4O_{12}$ |
|---|---|---|---|---|
| 1.751_$CaCo_3Ti_4O_{12}$ | 3.24_$CaFe_3Ti_4O_{12}$ | 1.245_$CoBr_2$ | 0.395_MnPtGa | 0.155_$CaMnGe_2O_6$ |
| 0.297_$NaCrGe_2O_6$ | 0.504_$NaCrSi_2O_6$ | 1.260_$NaMnGe_2O_6$ | 0.21_$PbNiO_3$ | 0.50_$MnTiO_3$ |
| 0.112_$FeBO_3$ | 0.113_$NiCO_3$ | 0.114_$CoCO_3$ | 0.115_$MnCO_3$ | 0.116_$FeCO_3$ |
| 0.416_$LaCrO_3$ | 0.463_$Co_3O_4$ | 0.108_$Mn_3Ir$ | 1.508_$Mn_2AlB_2$ | 0.72_$CaMnBi_2$ |
| 0.73_$SrMnBi_2$ | 0.611_$BaMnSb_2$ | 0.896_$NiCrO_4$ | 1.520_$NiSO_4$ | 1.521_$FeSO_4$ |
| 1.522_$CrVO_4$ | 1.523_$VPO_4$ | 1.558_$MnSn_2$ | 1.556_$FeSn_2$ | 1.557_$FeGe_2$ |
| 0.199_$Mn_3Sn$ | 0.279_$Mn_3As$ | 0.377_$Mn_3Ge$ | 0.19_$MnTiO_3$ | 1.304_$ZnMnO_3$ |
| 1.580_$NiTiO_3$ | 0.20_$MnTe_2$ | 1.18_$MnS_2$ | 0.150_$NiS_2$ | 1.305_$Mn_5Si_3$ |
| 0.598_$AlCr_2$ | 0.639_$Mn_2Au$ | 0.59_$Cr_2O_3$ | 0.65_$Fe_2O_3$-alpha | 1.287_$V_2O_3$ |
| 1.767_$Li_2CoCl_4$ | 0.15_$MnF_2$ | 0.36_$NiF_2$ | 0.178_$CoF_2$ | 0.163_$MnPS_3$ |
| 1.183_$FePS_3$ | 1.230_$NiPS_3$ | 1.264_$CoPS_3$ | 0.191_$BaCuF_4$ | 1.64_$BaNiF_4$ |
| 1.439_$BaCoF_4$ | 0.284_$KOsO_4$ | 0.285_$KRuO_4$ | 0.633_$KFeS_2$ | 0.636_$RbFeS_2$ |
| 0.637_$KFeSe_2$ | 0.638_$RbFeSe_2$ | 0.712_$VNb_3S_6$ | 3.26_$CoNb_3S_6$ | 1.0.8_$Ba_3MnNb_2O_9$ |
| 1.246_$CoCl_2$ | 1.247_$NiCl_2$ | 1.248_$NiBr_2$ | 1.244_$CrCl_3$ | 1.782_$FeBr_3$ |
| 1.786_$RuBr_3$ | 1.787_$RuCl_3$ | 1.733_$Cu_2MnGeS_4$ | 1.57_$CuMnO_2$ | 1.61_$MnWO_4$ |
| 1.194_$NiWO_4$ | 1.653_$FeWO_4$ | 1.742_$KNiAsO_4$ | 1.860_$KCoAsO_4$ | 0.1065_$MnGeN_2$ |
| 0.1066_$MnSiN_2$ | 0.180_$MnPSe_3$ | 1.210_$FePSe_3$ | 0.215_$BaNi_2P_2O_8$ | 1.256_$BaNi_2V_2O_8$ |
| 1.257_$BaNi_2As_2O_8$ | 0.334_$CoF_3$ | 0.335_$FeF_3$ | 0.434_$K_2ReI_6$ | 1.785_$K_2ReCl_6$ |
| 0.528_CrSb | 0.800_MnTe | 0.737_$LaBaMn_2O_6$ | 1.0.27_$Li_2MnTeO_6$ | 1.17_$CoV_2O_6$-alpha |
| 1.849_$Na_3Ni_2SbO_6$ | 1.23_$La_2CuO_4$ | 1.42_$La_2NiO_4$ | 1.249_$K_2NiF_4$ | 1.409_$NaMnO_2$ |
| 1.550_LiMnAs | 0.1018_$SrMnO_3$ | 0.761_$SrFe_2Se_2O$ | 0.762_$SrFe_2S_2O$ | 0.768_$SrMnSb_2$ |
| 0.834_$CrSbSe_3$ | 0.860_$Co_3Sn_2S_2$ | 0.987_$BaFe_2S_2O$ | 0.988_$BaFe_2Se_2O$ | 1.121_$NaFeSO_4F$ |
| 1.190_$YCr(BO_3)_2$ | 1.229_$BaMoP_2O_8$ | 1.37_VOCl | 1.802_CrSBr | 1.472_CaOFeS |
| 1.547_CsMnP | 1.186_$SrRu_2O_6$ | 1.593_BaCoSO | 1.805_$FeVMoO_7$ | 1.806_$CrVMoO_7$ |
| 1.854_$CoTeMoO_6$ | 0.1004_$CsO_2$ | 0.128_$FeSO_4F$ | 0.399_FeOOH | 0.433_$KMnF_3$ |
| 0.733_$AgRuO_3$ | 0.760_$FeOHSO_4$ | 0.798_$MnPd_2$ | 0.802_$CuFeS_2$ | 0.823_$Sr_2MnGaO_5$ |
| 1.136_$AgCrS_2$ | 1.219_$CuF_2$ | 1.26_$CsFe_2Se_3$ | 1.281_$YBaCuFeO_5$ | 1.308_$MnBi_2Te_4$ |
| 1.370_$Li_2CuO_2$ | 1.388_$La_2NiO_3F_2$ | 1.403_$La_2CoO_4$ | 1.404_$Sr_2CuO_2Cl_2$ | 1.418_$Cu_4O_3$ |
| 1.478_$CoTi_2O_5$ | 1.526_$LiCoF_4$ | 1.62_CuO | 1.642_$TlFeS_2$ | 1.65_$SrFeO_2$ |
| 1.678_CrN | 1.724_$Ba_2NiTeO_6$ | 1.735_$Li_2FeGeS_4$ | 1.803_$LiCrTe_2$ | 1.838_$LaMn_2Au_4$ |
| 1.840_$Sr_2Mn_2O_5$ | 1.841_$Sr_3Fe_2O_5Cl_2$ | 1.875_$KV_2Se_2O$ | 1.9_$Li_2VOSiO_4$ | |

TABLE II. The predicted magnetic structures for which the calculated total energy of the experimental magnetic structure is more than 5 meV/mag. atom higher than that of the computed lowest-energy configuration. For 1.456_$Sr_2CuO_2Cu_2S_2$, 0.374_$YNi_4Si$, 1.5_$YBa_2Cu_3O_{6+d}$, 1.397_$Cu_3Mg(OD)_6Br_2$, and 1.420_$YBa_2Cu_3O_6$, the optimized magnetic moments are nearly quenched. Each entry is designated by its standardized ID in the MAGNDATA repository followed by its chemical formula.

| Materials | | | | |
|---|---|---|---|---|
| 0.364_$SrCr_2As_2$ | 0.472_$LaMn_2Si_2$ | 1.252_$CaCo_2P_2$ | 1.458_$CsCo_2Se_2$ | 1.469_$YMn_2Si_2$ |
| 1.496_$YMn_2Ge_2$ | 1.462_$La_2CoPtO_6$ | 0.461_$CoRh_2O_4$ | 1.24_$ZnV_2O_4$ | 1.618_CoO |
| 0.275_$Mn_3AlN$ | 1.153_$Mn_3GaC$ | 0.501_$LiFe_2F_6$ | 0.1023_$Cr_2MoO_6$ | 1.0.14_$CsFeCl_3$ |
| 1.0.26_$RbCoBr_3$ | 1.0.40_$RbFeCl_3$ | 0.212_$Sr_2Mn_3As_2O_2$ | 1.110_$ScMn_6Ge_6$ | 1.631_$YMn_6Ge_6$ |
| 1.758_$CaMn_3V_4O_{12}$ | 1.846_$LaCu_3Fe_4O_{12}$ | 1.242_$FeBr_2$ | 1.154_$NaFeSi_2O_6$ | 1.169_$CaCoGe_2O_6$ |
| 0.109_$Mn_3Pt$ | 0.414_$AlFe_2B_2$ | 0.125_$MnGeO_3$ | 0.277_$MgMnO_3$ | 1.232_CuMnSb |
| 0.607_$RuO_2$ | 1.241_$FeCl_2$ | 2.116_$Na_3Co_2SbO_6$ | 2.23_$Sr_2CoO_2Ag_2Se_2$ | 2.24_$Ba_2CoO_2Ag_2Se_2$ |
| 0.401_$Sr_4Fe_4O_{11}$ | 0.1147_$CrSbS_3$ | 1.159_$Li_2Ni(WO_4)_2$ | 1.452_FeSn | 1.629_FeGe |
| 0.236_$CaFe_4Al_8$ | 0.266_$Na_2BaCo(VO_4)_2$ | 0.79_$CaIrO_3$ | 1.25_$KFe_3(OH)_6(SO_4)_2$ | 1.278_$Cu(NCS)_2$ |
| 1.345_$NaMnF_4$ | 1.346_$TlMnF_4$ | 1.702_$YBaCo_2O_5$ | 1.830_$FeBi_4S_7$ | 1.97_$Li_2MnO_3$ |
| 1.456_$Sr_2CuO_2Cu_2S_2$ | 0.374_$YNi_4Si$ | 1.449_$Li_2CuW_2O_8$ | 1.5_$YBa_2Cu_3O_6$+d | 1.397_$Cu_3Mg(OD)_6Br_2$ |
| 1.420_$YBa_2Cu_3O_6$ | | | | |